\documentclass[11pt]{article}
\usepackage{latexsym}
\usepackage{amsmath}
\usepackage{pifont}
\usepackage{wasysym}

 \csname
@addtoreset\endcsname{equation}{section}

\def\pe{\prime}
\def\3s{{s \choose 3}}
\def\4s{{s \choose 4}}
\def\5s{{s \choose 5}}
\def\6s{{s \choose 6}}
\def\12{\frac{1}{2}}
\def\fr{\frac}
\def\pr{\partial}
\def\prd{\partial \cdot}

\def\ra{\rightarrow}

\def\bec{\begin{center}}
\def\ec{\end{center}}
\def\a{\alpha}  

\def\b{\beta}

\def\d{\delta} 
\def\D{\Delta}
\def\e{\epsilon}

\def\vf{\varphi}

\def\l{\lambda}
\def\L{\Lambda}
\def\m{\mu}
\def\n{\nu}
\def\r{\rho}
\def\s{\sigma}

\def\th{\theta}

\def\h{\eta}

\def\cB{{\cal B}}

\def\cG{{\cal G}}
\def\cE{{\cal E}}
\def\g{\gamma}
\def\cJ{{\cal J}}
\def\cC{{\cal C}}
\def\cL{{\cal L}}
\def\cD{{\cal D}}
\def\cF{{\cal F}}

\def\cA{{\cal A}}
\def\cW{{\cal W}}

\def\cR{{\cal R}}
\def\cS{{\cal S}}

\def\cH{{\cal H}}

\def\cA{{\cal A}}

\def\ra{\rightarrow}

\newcommand{\eq}[1]{(\ref{#1})}

\def\be{\begin{equation}}
\def\ee{\end{equation}}
\def\bea{\begin{eqnarray}}
\def\eea{\end{eqnarray}}
\def\ba{\begin{array}}
\def\ea{\end{array}}

\def\ft#1#2{{\textstyle{{\scriptstyle #1}
\over {\scriptstyle #2}}}}

\def\scs#1{\section{\sc #1}}

\def\dsl{\not { \pr}}
\def\dsll{\not {\! \pr}}
\def\psisl{\not {\! \! \psi}}

\def\e{\epsilon}
\def\esl{\not {\! \epsilon}}

\def\ssl{\not {\! \cal S}}

\def\esl{\not {\! \epsilon}}
\def\ssl{\not {\! \cal S}}

\def\mp{\m_{\, \psi}}
\def\mpsl{\not {\!  \m_{\, \psi}}}

\def\ba{\begin{align}}
\def\ena{\end{align}}
\def\be{\begin{equation}}
\def\ee{\end{equation}}
\def\fr{\frac}
\def\ft{\footnote}

\def\a{\alpha}
\def\b{\beta}
\def\g{\gamma}
\def\G{\Gamma}
\def\d{\delta}
\def\D{\Delta}
\def\e{\epsilon}

\def\h{\eta}
\def\th{\theta}

\def\l{\lambda}
\def\L{\Lambda}
\def\m{\mu}
\def\n{\nu}

\def\r{\rho}

\def\s{\sigma}

\def\vf{\varphi}

\def\ps{\psi}

\def\cB{{\cal B}}
\def\cE{{\cal E}}
\def\cG{{\cal G}}
\def\cJ{{\cal J}}
\def\cK{{\cal K}}
\def\cC{{\cal C}}
\def\cL{{\cal L}}
\def\cD{{\cal D}}
\def\cF{{\cal F}}

\def\cA{{\cal A}}
\def\cW{{\cal W}}

\def\cR{{\cal R}}
\def\cS{{\cal S}}

\def\cH{{\cal H}}

\def\cA{{\cal A}}

\def\psb{{\bar\psi}}

\def\dsl{\not { \pr}}
\def\dsll{\not {\! \pr}}

\def\psisl{\not {\! \! \psi}}

\def\esl{\not {\! \epsilon}}
\def\ssl{\not {\! \cal S}}

\def\rsll{\not {\! \! \cal R}}

\def\esl{\not {\! \epsilon}}
\def\ssl{\not {\! \cal S}}

\def\gsl{\not {\!  \Gamma}}

\def\gz0{\gamma^{0}}

\def\pe{\prime}
\def\eq{\equiv}

\def\ra{\rightarrow}

\def\pr{\partial}

\def\3s{{s \choose 3}}
\def\4s{{s \choose 4}}
\def\5s{{s \choose 5}}
\def\6s{{s \choose 6}}
\def\12{\frac{1}{2}}

\def\prd{\pr \cdot}

\begin{document}

\begin{flushright}
\vskip 8pt {\today}
\end{flushright}

\vspace{5pt}

\begin{center}

%%%%%%%%%%%%%%%%%%%%%%%%%%%%%%%%%%%%%%%%%%%%%%%%%%%%%%%%%%%%%%%%%%%%

{\Large\sc Geometric Lagrangians for massive higher-spin fields}\\

%%%%%%%%%%%%%%%%%%%%%%%%%%%%%%%%%%%%%%%%%%%%%%%%%%%%%%%%%%%%%%%%%%%%

\vspace{25pt}
{\sc D.~Francia}\\[15pt]

\it\sl\small Department of Fundamental Physics\\ Chalmers
University of Technology \\ S-412\ 96 \ G\"oteborg \ SWEDEN
\\ e-mail: {\small \it francia@chalmers.se}\vspace{10pt}

%%%%%%%%%%%%%%%%%%%%%%%%%%%%%%%%%%%%%%%%%%%%%%%%%%%%%%%%%%%%%%%%%%%
\vspace{40pt} {\sc\large Abstract}\end{center}

{Lagrangians for massive, unconstrained, higher-spin bosons and fermions are proposed.
The idea is to modify the geometric, 
gauge invariant Lagrangians describing the corresponding massless theories
by the addition of  suitable quadratic polynomials. These
polynomials provide generalisations of the Fierz-Pauli mass term
containing all possible traces of the basic field. 
No auxiliary fields are needed.}

 \setcounter{page}{1}

\pagebreak

\tableofcontents

\newpage

%%%%%%%%%%%%%%%%%%%%%%%%%%%%%%%%%%%%%%%%%%%%%%%%%%%%%%%%%%%%%%%%%%%%%

%%
\scs{Introduction}\label{sec:Intro}

 The central object in the theories of massless 
spin-$1$ and spin-$2$ fields is the curvature. It contains the informations needed 
for the classical description of the dynamics,
and reflects its geometrical meaning.
For higher-spin gauge fields\ft{For reviews see \cite{review}. 
An overview on the subject can also be found in
the Proceedings of the First Solvay Workshop on Higher-Spin Gauge
Theories \cite{solvay04}, available on the website 
http://www.ulb.ac.be/sciences/ptm/pmif/Solvay1proc.pdf .} 
such a description of the dynamics
based on curvatures is 
missing\ft{We are referring here to a ``metric-like''
formulation, generalisation of the
corresponding formulation for Gravity
where the basic role is played by the metric
tensor $g_{\, \m \n}$. 
Vasiliev's construction of non-linear
equations of motion for higher-spin
gauge fields \cite{vas1} represents a generalisation 
of the ``frame-like'' formulation of 
Einstein's theory.}
at the full interacting level, but it can exhibited
at least for the linear theory, where it
already displays a sensible amount of non-trivial features
\cite{fs1, Bekaert:2002dt, fs2, nonsym1, hull, Bekaert:2003zq, 
fs3, bb06, fsRev, FMS, Manvelyan:2007hv, joan2}.

 In this work we would like to show that the geometric Lagrangians
proposed  in \cite{fs1, fs2, FMS}
admit relatively simple quadratic
deformations, so as to provide a consistent description
of the corresponding massive theory.

 In the case of symmetric, rank-$s$ Lorentz  tensors  $\vf_{\, \m_1 \, \dots \, \m_s}$
and spinor-tensors $\psi_{\, \m_1 \, \dots \, \m_s}^{\, a}$,  
to which we shall restrict our attention in this work, 
the conditions to be met by these fields in order to describe the free propagation of  massive, 
irreducible representations of the Poincar\'e group
are contained in  the Fierz systems \cite{Fierz}:
\begin{align} \label{fierzsyst}
&(\Box \, - \, m^2) \, \vf_{\, \m_1 \, \dots \, \m_s} \, = \, 0\, ,& & &
&(i\,\g^{\, \a}\,\pr_{\, \a} \, - \, m) \, \psi_{\, \m_1 \, \dots \, \m_s}^{\, a} \,  = \,  0 \, ,\nonumber\\
&\pr^{\, \a}  \vf_{\, \a \,  \m_2\, \dots \, \m_s} \, = \,  0 \, ,& & &
&\pr^{\, \a}  \psi_{\, \a \,  \m_2\, \dots \, \m_s}^{\, a} \,  = \,  0\, , \\
&\vf^{\, \a}_{\ \ \  \a \, \m_3 \, \dots \, \m_s} \,   = \, 0 \, , & & &
&\g^{\, \a}\, \psi_{\, \a \, \m_2 \, \dots \, \m_s}^{\, a} \, = \, 0 \, . \nonumber
\end{align}

The quest for a Lagrangian description of these systems
has been a basic field-theoretical issue since Fierz and Pauli  proposed it in 
\cite{fierzpauli}, and several approaches and solutions are known up to now, 
both for flat and, more generally, for maximally symmetric spaces, 
\cite{massiveSH, massiveKK, ADY87, Aulakh:1986cb,
Deser, Zinoviev, deMedeiros, metsaev,
Hallowell, massiveBRST}
(for other results on massive higher-spins see
\cite{Weinberg}). 
Typically in these solutions auxiliary fields are present for spin $ s \geq \fr{5}{2}$
and, because of the interplay among the various fields, 
the proposed Lagrangians in general do not look like simple 
quadratic deformations of the corresponding massless ones. 

 This is to be contrasted with what happens for  
spin $s \leq 2$. Indeed, the Maxwell Lagrangian 
supplemented with a mass term
\be
\cL \, = \, - \, \fr{1}{4} \, F_{\m \n} \, F^{\m \n} \, 
- \, \12 \, m^{\, 2} \, A_{\m} \, A^{\m} \, ,
\ee
gives equations of motion that are easily shown to
imply $\pr_{\m} \, A^{\, \m} \, = \, 0$, 
the only condition needed to recover in this case 
the Fierz system. 
 A bit less direct, and more instructive for us, 
is the corresponding result for the massive graviton 
\cite{Boulware:1973my}.
Consider the linearised Einstein-Hilbert Lagrangian,
deformed by the introduction of a so-called
Fierz-Pauli mass term \cite{fierzpauli}:
\be \label{FPspin2}
\cL \, = \, \12 \, h^{\, \m \n}\, \{\cR_{\m \n} \, - \, \12 \, \h_{\m \n} \, \cR \, - \, m^{\, 2} \, 
(h_{\m \n} \, - \, \h_{\m \n}h^{\, \a}_{\, \ \a})\} \, ,
\ee
where $\cR_{\m \n}$ and $\cR$ indicate the linearised
Ricci tensor and Ricci scalar respectively.
The key point is that the divergence of the corresponding
equation of motion, because of the Bianchi identity 
$\pr^{\, \a} \{\cR_{\a \m}  -  \12 \, \h_{\a \m} \, \cR\} \,\eq \, 0$,
implies an on-shell constraint on the mass term of the form
\be \label{fpconstr}
\pr^{\, \a} \,  h_{\, \a \, \m} \, - \, \pr_{\, \m} \, h^{\, \a}_{\, \ \a} \, = \, 0 \, ,
\ee
which in the following 
we shall refer to as \emph{the Fierz-Pauli constraint}.
It is then simple to see that the double divergence of the mass term
is proportional to the Ricci scalar, so that the divergence of (\ref{fpconstr})
implies
\be
\cR \, = \, 0 \, .
\ee
The trace of the resulting equation
\be
\cR_{\m \n} \, - \, m^{\, 2} \, 
(h_{\m \n} \, - \, \h_{\m \n} \, h^{\, \a}_{\, \ \a}) \, = \, 0 \, ,
\ee
implies in turn $h^{\, \a}_{\, \ \a} = 0$, 
and then the full Fierz system (\ref{fierzsyst}) can be recovered noticing
that, under the conditions of vanishing  trace and vanishing divergence of 
$h_{\m  \n}$, the Ricci tensor reduces to $\Box \, h_{\m \n}$.
In \cite{Boulware:1973my, VanNieuw} it was shown 
that the Fierz-Pauli mass term in (\ref{FPspin2})
defines the  \emph{unique} quadratic deformation
of the linearised Einstein-Hilbert action free
of tachyons or ghosts.
 
 To summarise, both for spin $1$ and spin $2$ the key idea is 
to exploit the Bianchi identities of the ``massless sector'' of the 
equations of motion in order to 
derive the on-shell conditions (\ref{fpconstr}) from the divergence of 
a properly chosen
mass term. These conditions reveal 
necessary and sufficient to recover the Fierz system (\ref{fierzsyst}), 
and in this sense it is clear 
that such an approach cannot be tried, without modifications, in the absence
of a divergenceless Einstein tensor. 

 This is the reason why
this idea cannot work in the constrained description
of massless higher-spin bosons given by Fronsdal in 
\cite{fronsdal}, which is known, on the other hand,  to correctly 
describe the free propagation of gauge fields of integer spin.
In that framework indeed the analogue of the 
Ricci tensor, defined as 
\be  \label{fronsdalT}
\cF_{\m_1 \, \dots \, \m_s} \, \equiv \, \Box \, \vf_{\m_1 \, \dots \m_s}\, -  \, 
\pr_{\m_1} \, \pr^{\, \a} \,  \vf_{\a \, \m_2 \, \dots \, \m_s} \, + \, \dots \, + \, 
\pr_{\m_1}\pr_{\m_2}\, \vf_{\, \ \a \,\m_3 \, \dots \m_s}^{\, \a}
+ \, \dots \, ,
\ee
where the dots indicate symmetrization over the set of $\m$-indices,
is used to build an Einstein tensor of the form
\be \label{fronsdalE}
\cE_{\m_1 \, \dots \m_s} \, = \, \cF_{\m_1 \, \dots \, \m_s} \, - \, \12 \, 
(\h_{\, \m_1 \m_2} \, \cF^{\, \a}_{\ \ \a \, \m_3 \, \dots \m_s} \, + \, \dots) ,
\ee
whose divergence gives identically
\be \label{fronsdalD}
\pr^{\, \a} \, \cE_{\, \a \, \m_2 \, \dots \m_s} \, \eq \, - \, \fr{3}{2} \, 
(\pr_{\, \m_2} \pr_{\, \m_3} \pr_{\, \m_4} \, \vf^{\, \a \, \b}_{\ \ \ \a \, \b  \, \m_5 \, \dots \m_s} \, + \, 
\dots \,) \,  - \, \12 \, (\h_{\, \m_2 \m_3} \, \pr^{\, \a} \cF^{\, \b}_{\, \ \a \, \b \, \m_4 \, \dots \m_s} \, + \, \dots) \, .
\ee
It is then clear that, even under the condition 
of vanishing double trace assumed in that context,
\be \label{fronsdalDT}
\vf^{\, \a \, \b}_{\ \ \ \a \, \b  \, \m_5 \, \dots \m_s} \, \eq \, 0 \, ,
\ee
the divergence of $\cE_{\m_1 \, \dots \m_s}$ is not zero, but still retains a trace
part. 

In strict analogy, in the Fang-Fronsdal theory of massless fermionic
fields \cite{ffronsdal} a constraint is assumed  on the triple gamma-trace of the basic
field\ft{From now on we drop the spinor index $a$ in $\ps^{\, a}_{\, \m_1 \, \dots \, \m_s}$.}
\be \label{ffronsdalDT}
\g^{\, \a} \, \g^{\, \b} \, \g^{\, \g} \, \ps_{\, \a \, \b \, \g \, \m_4 \, \dots \, \m_s} \, 
\eq \, 0 \, ,
\ee
and the dynamics is described in terms of a generalisation 
of the Dirac-Rarita-Schwinger tensor, having the form
\be \label{ffronsdalT}
\cS_{\, \m_1 \, \dots \, \m_s} \, \eq \, i\, 
\{\g^{\, \a}\,\pr_{\, \a} \, \, \psi_{\, \m_1 \, \dots \, \m_s}
\, - \, (\pr_{\, \m_1} \, \g^{\, \a}\, \psi_{\, \a \, \m_2 \, \dots \, \m_s} \, 
+ \, \dots \, )\}\, .
\ee
Similarly to (\ref{fronsdalD}), the divergence
of the corresponding
Einstein tensor
\be \label{ffronsdalE}
\cG_{\m_1 \, \dots \m_s} \, \eq \, \cS_{\m_1 \, \dots \, \m_s} \, - \, \12 \, 
(\g_{\, \m_1}\, \g^{\, \a} \, \cS_{\, \a \, \m_2 \, \dots \, \m_s}\, + \, \dots) \, 
- \, \12 \, 
(\h_{\, \m_1 \m_2} \, \cS^{\, \a}_{\ \ \a \, \m_3 \, \dots \m_s} \, + \, \dots) ,
\ee
does not vanish even if (\ref{ffronsdalDT})
is imposed, but still retains
$\g$-trace parts, according to the general identity
\be \label{ffronsdalD}
\begin{split}
\pr^{\, \a} \, \cG_{\, \a \, \m_2 \, \dots \m_s} \, = &
- \, \12 \, (\h_{\, \m_2 \m_3} \, \pr^{\, \a} \cS^{\, \b}_{\ \ \a \, \b \, \m_4 \, \dots \m_s} \, + \, \dots)
\, - \, \12 \, (\g_{\, \m_2}\, \pr^{\, \a} \, \g^{\, \b} \, 
\cS_{\ \ \a\, \b \, \m_3 \, \dots \, \m_s}\, + \, \dots) \, \\
& + \, i \, 
\, (\pr_{\, \m_2} \,  \pr_{\, \m_3} \, 
\g^{\, \a} \, \g^{\, \b} \, \g^{\, \g} \, \ps_{\, \a \, \b \, \g \, \m_4 \, \dots \, \m_s}  \, + \, \dots)\, .
\end{split}
\ee

On the other hand, 
the Lagrangians proposed in \cite{fs1, fs2, FMS} are
based on \emph{identically divergenceless} Einstein tensors, 
for which it is possible in principle to try the extension 
of the spin-$1$ and spin-$2$ results. This construction is the object of this work.

The bosonic and the fermionic cases
are presented separately, in Section $2$ and Section $3$ respectively.
In particular, Section \ref{section2.1} contains a review of the construction 
of geometric Lagrangians for bosons proposed in \cite{fs1, fs2, FMS},
with focus on the main conceptual ideas behind it.
Accordingly, in Section \ref{section2.1.1} the definition of 
higher-spin curvatures following de Wit and Freedman \cite{dewfr}
is recalled, and in Section \ref{section2.1.2} it is discussed 
how to derive from these curvatures candidate ``Ricci tensors''
$\cA_{\, \m_1 \, \dots \, \m_s}$,
supposed to define basic equations of motion in the form 
$\cA_{\, \m_1 \, \dots \, \m_s} = 0$. In Section \ref{section2.1.3}
the problem of constructing \emph{identically
divergenceless} Einstein tensors,
and corresponding gauge-invariant Lagrangians, is reviewed.

 The main issue to be stressed, following \cite{FMS}, is that
this geometric program has \emph{infinitely many solutions}.
Nonetheless, it was shown in \cite{FMS} that problems arise 
when a coupling with an external current is turned on, and 
a closer analysis proves that 
 \emph{only one} geometric theory
passes the test of defining consistent current exchanges.
This theory also results to be the only one for which a 
clear local counterpart exists,
described by the local Lagrangians proposed in 
\cite{fs3, fsRev}.

 For both bosons and fermions the geometric Lagrangians 
can not be standard second-order or first-order ones
since,  for  $s \geq \fr{5}{2}$, the curvatures 
are  higher-derivative tensors, 
and the corresponding  equations unavoidably
contain non-localities, if one wishes to preserve at least 
formally the dimensions of canonical, relativistic 
wave operators. Because of this,  one of the key issues is to prove their compatibility with 
the (Fang-)Fronsdal theory,
synthetically defined by (\ref{fronsdalT}) (\ref{fronsdalE}) (\ref{fronsdalD}) 
(\ref{fronsdalDT}) for bosons, and (\ref{ffronsdalDT})
(\ref{ffronsdalT}) (\ref{ffronsdalE}) (\ref{ffronsdalD}) 
for fermions, together with the conditions
that the abelian gauge transformations of the fields
\begin{align} \label{fierz}
&\d \, \vf_{\m_1 \, \dots \m_s} \, = \, \pr_{\, \m_1} \,
\L_{\m_2 \, \dots \m_s} \, + \, \dots \, , & & &
&\d \, \psi_{\m_1 \, \dots \m_s} \, = \, \pr_{\, \m_1} \,
\e_{\m_2 \, \dots \m_s} \, + \, \dots \, , 
\end{align}
be understood in terms of ($\g$-)\emph{traceless} gauge parameters
\begin{align} \label{fronsdalTG}
&\L^{\, \a}_{\ \ \a \, \m_3 \, \dots \m_{s - 1}} \, \eq \, 0 \, , & & &
&\g^{\, \a} \, \e_{\, \a \, \m_2 \, \dots \m_{s - 1}} \, \eq \, 0 \, ,
\end{align}
as needed for the gauge invariance of the equations 
\begin{align} \label{fanfro}
&\cF_{\m_1 \, \dots \m_s} \, = \, 0\, ,& & & 
&\cS_{\m_1 \, \dots \m_s} \, = \, 0\, ,
\end{align}
respectively. This is achieved by showing that the 
partial gauge-fixing needed to remove all
the singularities from the geometric equations of motion
simultaneously reduces those equations to the local, 
constrained form (\ref{fanfro}).

 The counterpart of the singularities of the geometric theory
on the side of the local, unconstrained Lagrangians
proposed in \cite{fs3, fsRev}, and briefly recalled
in Section \ref{section2.1.3}, can be recognised in the presence of 
higher-derivative terms in the kinetic operator of a
``compensator'' field, $\a_{\m_1 \, \dots \, \m_{s - 3}}$, introduced 
in that context in order to eliminate the non-localities of the
irreducible formulation. Since this field represents a pure-gauge
contribution to the equations of motion, needed to guarantee
invariance under a wider symmetry than the constrained one
of (\ref{fronsdalTG}), the corresponding higher-derivative
terms, although they could in general  be the source
of problems, both at the classical and at the quantum level, 
should not interfere in this case with the physical content 
of the theory itself. 

 Nonetheless, on the side of the interesting quest 
for an ordinary-derivative formulation
of the same dynamics, an important 
result  was recently achieved in  
\cite{Buchbinder:2007ak}, where a 
two-derivative formulation for unconstrained  systems of
arbitrary-spin fields was 
proposed, in a rather economical description
involving for any spin only a limited (and fixed)
number of fields.\ft{The first description
of unconstrained, ordinary-derivative, higher-spin dynamics was 
given in \cite{pt}. The BRST construction proposed in those
works involves an unconstrained spin-$s$
symmetric tensor together with additional fields, whose number grows
proportionally to $s$.}.
Another relevant feature of this work is the stress
put on the close relationship with the triplet systems 
of Open String Field Theory \cite{oldtriplet} previously
investigated from this point of view in
\cite{fs2, st}\ft{See also \cite{bonelli} for related contributions, and 
\cite{miriancubic, mirianvertex} for recent developments.}.
 
 Here we propose a similar result, for the bosonic case, showing
in Section \ref{section2.1.4} that the same dynamical content of the 
local Lagrangians
of \cite{fs3, fsRev} can also be expressed in terms of an 
ordinary-derivative Lagrangian,  
at the price of enlarging  the field
content of the theory from the minimal
set of three fields of \cite{fs3, fsRev}, to a slightly bigger one 
involving five fields altogether, for any spin.
This result, with respect
to the one presented in \cite{Buchbinder:2007ak}, 
looks more closely related to the underlying 
geometric description of \cite{fs1, fs2, FMS}, 
which, in a sense, should provide its 
ultimate meaning.

 In the discussion of fermions, in Section \ref{section3},
more space is devoted to the analysis
of the fermionic geometry, which 
is less straightforward than the 
corresponding bosonic one, and for which less details
were given in \cite{fs1, fs2}, where the construction 
of gauge-invariant, non-local, kinetic tensors  for fermions
was substantially deduced from the knowledge of the corresponding
bosonic ones.
The analysis of the fermionic geometry 
performed in Sections \ref{section3.1.1} and  \ref{section3.1.2} 
provides first of all the explicit expression of the kinetic tensors of
\cite{fs1, fs2} in terms of curvatures, and shows the 
existence of a wider range of possibilities for the
construction of basic candidate ``Dirac'' tensors, whose
meaning is clarified in Section \ref{section3.1.3}, under the
requirement that a Lagrangian derivation of the 
postulated equations of motion be possible.

 The structure
of the mass deformation and the  proof
that the Fierz systems (\ref{fierzsyst}) are actually recovered on-shell
are given in Section \ref{section2.2} 
for bosons and in Section \ref{section3.2} 
for fermions.  The main result will be that the 
Fierz-Pauli mass term (\ref{FPspin2})
is actually the beginning of a sequence involving all possible ($\g$-)traces 
of the field, according to eqs. (\ref{mass}) and (\ref{massferm}), 
whereas the guiding principle in the construction will be to recognise the need
for the implementation of the Fierz-Pauli constraint (\ref{fpconstr})
(and of its fermionic counterpart) for any spin.
 
 To give an idea of the outcome, the generalised Fierz-Pauli mass term
for spin $4$ has the form\ft{With $M_{\, \vf}$ we indicate here  the linear
combination of traces of $\vf$ entering the Lagrangian in the 
schematic form $\cL = \12 \vf \, \{\cE_{\, \vf} \, - \, m^{\, 2} \, M_{\, \vf}\}$,
where $\cE_{\, \vf}$ is the Einstein tensor of the corresponding
massless theory. Consequently, the equations of motion will appear in 
the form $\cE_{\, \vf} \, - \, m^{\, 2} \, M_{\, \vf} = 0$. Similarly
for the fermionic mass-term, $M_{\, \psi}$.}
\be  \nonumber
M_{\, \vf} \, = \,  \vf_{\, \m_1 \dots \m_4} \, - \, (\h_{\, \m_1 \m_2} \, 
\vf^{\, \a}_{\ \ \a \, \m_3 \m_4} \, + \, \dots\,) \, 
- \, (\h_{\, \m_1 \m_2} \, \h_{\, \m_3 \m_4} \, + \, \dots \, )\, 
\vf^{\, \a \,  \b}_{\ \ \a \, \b} \, ,
\ee
whereas the corresponding result for spin $\fr{7}{2}$ is
\be \nonumber
M_{\, \psi} \, = \, \psi_{\, \m_1 \m_2 \m_3} \, - \,
(\g_{\, \m_1}\, \g^{\, \a} \, \psi_{\, \a \m_2 \m_4}\, + \, \dots) \,
- \,   (\h_{\, \m_1 \m_2} \, 
\psi^{\, \a}_{\ \ \a \, \m_3} \, + \, \dots\,) \, 
- \, (\g_{\, \m_1} \, \h_{\, \m_2 \m_3}  \, + \, \dots \, )\, 
\g^{\, \a} \, \psi^{\, \b}_{\ \ \a \, \b} \, .
\ee

 As a matter of principle, all of the infinitely many
geometric Lagrangians describing the same free dynamics, 
being built from divergenceless Einstein tensors, are
equivalently amenable to the quadratic deformation by means 
of the generalised Fierz-Pauli mass terms given in  
(\ref{mass}) and (\ref{massferm}). In this sense, 
an issue of uniqueness is present at the massive level
as well, as discussed in Section $4$,
and the analysis of the current exchange 
in this case is not sufficient to provide
a selection principle, as showed  
in the same Section by the example of spin $4$.

 Nonetheless, we believe that a direct relation 
exist between  the massive theories proposed in this work and the 
geometric theory described in \cite{FMS}, uniquely selected by the requirement
of consistency with the coupling to an external source.
To support this hypothesis, in Section \ref{section4}
we make an analysis of the 
form of the geometric solution introduced
in \cite{FMS}, and use the corresponding result 
to perform explicit computations of
the quantities of interest. Still, we were not able to give
a complete proof of this conjecture, which so far
is only supported by checks up to spin $11$. 

\begin{center}
\cancer
\end{center}
  
 In this work we shall mainly resort to a compact notation 
introduced in \cite{fs1, fs2}, in which
all symmetrized indices are left implicit. Traces will be indicated
by ``primes'' ($h^{\, \a}_{\ \ \a} \, \ra \, h^{\, \pe}$)
and divergences by the symbol ``$\prd$'' 
($\pr^{\, \a}\, h_{\, \a \m} \, \ra \, \prd h$), so that,
for example, the Fierz-Pauli constraint (\ref{fpconstr})
could be written in the index -free form
\be
\prd h \, - \, \pr \, h^{\, \pe}\,  = \, 0 \, .
\ee
The precise definitions 
and the corresponding computational rules needed in order
to take the combinatorics into account  are given in 
Appendix \ref{A}. 
 
 An exception is made when there will be the need 
to describe more than one group of symmetrized indices. 
In particular, since the de Wit-Freedman connections,
to be introduced in Section \ref{section2.1.1}, are rank-$(m + s)$
tensors where two group of $m$ and $s$ indices are separately
symmetrized, we found it more appropriate to
use a kind of ``mixed-symmetric'' notation. 
The presence of different groups of symmetrized indices will then be
\emph{explicitly} displayed   
by choosing the same letter for each index within a symmetric group.
So for example the rank-$(2+2)$ tensor indicated with $\pr_{\n} \, \pr_{\m}\vf_{\m \, \n}$
is to be understood as follows:
\be
\pr_{\n} \, \pr_{\m}\vf_{\m \, \n} \, \ra \, 
\pr_{\n_1} \, \pr_{\m_1}\vf_{\m_2 \, \n_2} \, + \, 
\pr_{\n_2} \, \pr_{\m_1}\vf_{\m_2 \, \n_1} \, + \, 
\pr_{\n_1} \, \pr_{\m_2}\vf_{\m_1 \, \n_2} \, + \, 
\pr_{\n_2} \, \pr_{\m_2}\vf_{\m_1 \, \n_1} \, ,
\ee
whereas in the general case of two groups of symmetrized indices, 
we shall use the shortcut notation
\be
\vf_{\, \m_1 \, \dots \, \m_m, \, \r_1 \, \dots \, \r_s} \ \ \ra \ \ 
\vf_{\, \m_m, \, \r_s}\, .
\ee
The rules of symmetric
calculus listed in Appendix \ref{A} apply in this notation separately
for each set of symmetric indices.

%%%%%%%%%%%%%%%%%%%%%%%%%%%%%%%%%%%%%%%%%%%%%%%%%%%%%%%%%%%%%%%%%%%%%

%%%%%%%%%%%%%%%%%%%%%%%%%%%%%%%%%%%%%%%%%%%%%%%%%%%%%%%%%%%%%%%%%%%%%

%%
\scs{Bosons}\label{bosons}
 \subsection{Geometry for higher-spin bosons} \label{section2.1}
 \subsubsection{Bosonic curvatures} \label{section2.1.1}

The starting point of the full construction is  the 
definition of \emph{higher-spin curvatures} given 
by de Wit and Freedman in \cite{dewfr} (see also 
\cite{damour}) and reviewed in 
\cite{fs1, fs2, bb06}.  These curvatures 
are the top elements of a hierarchy of generalised
``Christoffel connections'' 
$\G^{(m)}_{\, \m_1 \, \dots  \, \m_m,\,  \n_1 \, \dots \, \n_s}$
built from derivatives of the gauge field.
Roughly speaking, the rationale behind the construction
is to choose a linear combination of multiple gradients of $\vf$ 
($m$ gradients, for the $m$-th connection 
$\G^{(m)}_{\, \m_1 \, \dots \, \m_m,\, \n_1 \, \dots \, \n_s}$) such that,
under the transformation $\d \vf \, = \, \pr \, \L$, the 
gauge variation of the connections themselves become simpler and simpler, 
with increasing $m$. Specifically, in the mixed-symmetric notation,
the $m$-th connection is
\be \label{gammabose}
\G^{(m)}_{\, \m_{m},\, \n_{s}} =   \ \sum_{k=0}^{m}\fr{(-1)^{k}}{
\left(
{{m} \atop {k}}
\right)
}\ 
\pr^{\, m-k}_{\, \m}\pr^{\, k}_{\, \n} \, \vf_{\, \m_{k} \, \n_{s-k}}\, , 
\ee
whose gauge transformation is
\be \label{bosetrans}
\d\, \G^{(m)}_{\, \m_{m},\, \n_{s}} \, = \,(-1)^m \, (m+1)\, 
\pr^{\, m+1}_{\, \n} \, \L_{\, \m_{m},\, \n_{s-m-1}}\, . 
\ee

After $s$ steps, the resulting 
rank-$(s + s)$ tensor
is identically gauge invariant, and it is called
a (linearised) curvature.
For spin $1$ the outcome is simply the Maxwell
field strenght
\be
\cR_{\, \m, \n} \, = \, \pr_{\, \m} \, \vf_{\, \n} \, - \, \pr_{\, \n} \, \vf_{\, \m} \, ,
\ee
whereas for spin $2$ a linear combination of the usual
linearised Riemann tensors is obtained, that in terms
of the rank-$2$ field can be simply expressed as
\be
\cR_{\, \m \m,\, \n \n} \, = \, \pr^{\, 2}_{\, \m} \, \vf_{\n \, \n} \, - 
\12 \, \pr_{\, \n} \, \pr_{\, \m}\vf_{\m \, \n} \, + \,  \pr^{\, 2}_{\, \n} \, \vf_{\m\, \m} \, ,
\ee
where, as indicated in the Introduction, symmetrization is here assumed separately within
the $\m$-group and the $\n$-group of indices. In addition
$\cR_{\, \m \m,\, \n \n}$ is symmetric 
under the exchange of the two pairs, and cyclic. 
The same kind of symmetries are displayed in the generalisation 
to higher-spins of this pattern, so that for example
for spin $3$ the resulting curvature 
\be
\cR_{\, \m \m \m, \, \n \n \n} \, = \, \pr^{\, 3}_{\, \m} \, \vf_{\n \, \n \, \n} \, - 
\fr{1}{3} \, \pr^{\, 2}_{\, \m} \, \pr_{\, \n}\, \vf_{\m \, \n \, \n} \, + \,  
\fr{1}{3} \, \pr_{\, \m} \, \pr^{\, 2}_{\, \n}\, \vf_{\m \, \m \, \n} \, - \, 
\pr^{\, 3}_{\, \n} \, \vf_{\m\, \m \, \m} \, ,
\ee
is such that $\cR_{\, \m \m \m, \, \n \n \n} \, = \, - \cR_{\n \n \n, \, \m \m \m}$,
while in general for spin $s$ it can be checked that the tensors
\be \label{curvature}
\cR_{\, \m_s, \,  \n_s} \, 
= \, \sum_{k=0}^{s}\fr{(-1)^{k}}{
\left(
{{s} \atop {k}}
\right)
}\ 
\pr^{s-k}_{\, \m}\pr^{k}_{\, \n} \vf_{\, \m_k,\,\n_{s-k}} \, , 
\ee
defining proper generalisations of the linearised Riemann curvature, 
satisfy the constraints
\begin{align} %\label{onshellid}
&\cR_{\, \m_s, \,  \n_s} \, = \, (-)^{\, s} \, \cR_{\, \n_s, \,  \m_s} \, , &&&&&&
\cR_{\, \m_{s-1} \, \n, \,  \n_s} \, =\, 0 ,& 
\end{align}
the latter being the expression of the generalised cyclic 
identity\ft{Other definitions of ``curvature'' tensors, displaying
different symmetry properties, are possible. A curl on each index of
$\vf_{\, \m_1 \, \dots \, \m_s}$ defines a gauge-invariant, rank-$(s + s)$
tensor $\tilde{\cR}_{\, [\m_1 \n_1], \, \dots ,\, [\m_s \n_s]}$ antisymmetric
in each pair $(\m_j \n_j)$, more closely resembling the symmetry of the
Riemann tensor of gravity. This is the preferred choice
in the works \cite{Bekaert:2002dt, nonsym1}. Similarly, identical gauge invariance can 
also be reached taking a curl on one index of the 
$\G^{\, [s-1]}$ connection. All these possibilities can be shown to 
be related by linear combinations, and in this sense none can be 
considered \emph{a priori} preferable \cite{damour}. }.
Finally, generalised uncontracted Bianchi identities hold
for the full set of connections:
\be
\begin{split}
&\pr_{\, \l} \, \{\G^{(m)}_{\, \hat{\m}\, \m_{m-1}, \, \hat{\n}\, \n_{s-1}} \, - \, 
\G^{(m)}_{\, \hat{\n}\, \m_{m-1}, \, \hat{\m}\, \n_{s-1}} \} \, + \, \\
&\pr_{\, \hat{\n}} \, \{\G^{(m)}_{\, \l\, \m_{m-1}, \, \hat{\n}\, \n_{s-1}} \, - \, 
\G^{(m)}_{\, \hat{\m}\, \m_{m-1}, \, \l \, \n_{s-1}} \} \, + \, \\
&\pr_{\, \hat{\m}} \, \{\G^{(m)}_{\, \hat{\n} \, \m_{m-1}, \, \l \, \n_{s-1}} \, - \, 
\G^{(m)}_{\, \l \, \m_{m-1}, \, \hat{\n} \, \n_{s-1}} \} \, \eq \, 0 \, . 
\end{split}
\ee

  \subsubsection{Generalised Ricci tensors} \label{section2.1.2}

 For spin $1$ and spin $2$ the basic 
kinetic tensors (or ``Ricci'' tensors, with some abuse of terminology),
needed to define proper equations of motion, are obtained 
taking one divergence and one trace of the 
corresponding curvatures, respectively.
For integer spin $s$, starting from (\ref{curvature}),
the simplest possibility in order to
define a tensor with the same symmetries of the
gauge potential $\vf$ is to saturate all indices
belonging to the same group taking traces, while also taking 
one divergence for the leftover index for odd spins.
In all cases, however, in order to restore the dimension of a relativistic
wave operator, it is necessary to act with inverse powers
of the D'Alembertian operator, thus introducing
possible non-localities in the theory. For instance, the simplest choices
for candidate Ricci tensors for spin $3$ and $4$ 
are\ft{The notation $ \cF_{\, 2}$ refers to the fact that
for spin $1$ and spin $2$ the corresponding tensor is simply 
the standard Fronsdal tensor (\ref{fronsdalT}), indicated with $\cF \eq \cF_{\, 1}$.}
\be \label{kinetic3,4}
 \cF_{\, 2} \, \eq \,
     \begin{cases}
    \fr{1}{\Box} \, \prd \, \cR^{\, \pe} \, & \, s \, = \, 3 \, , \\
    \fr{1}{\Box} \, \cR^{\, \pe \pe} \, & \, s \, = \, 4 \, .
    \end{cases}
 \ee
On the other hand, once it is recognised that from the chosen viewpoint 
non-localities are unavoidable, there is no reason in principle to
discard the possibility that other tensorial structures could contribute. 
These must be selected taking into account possible identities
between apparently independent tensors.
For the case of $s = 3$, for example, out of the five possibilities 
\begin{align} \label{spinrstruct}
&\fr{\pr^{\, 2}}{\Box} \, \cF_{\, 2}^{\, \pe},  &
&\fr{\pr^{\, }}{\Box} \, \prd \cF_{\, 2},&
&\fr{\pr^{\, 3}}{\Box^{\, 2}} \, \prd \cF_{\, 2}^{\, \pe},&
&\fr{\pr^{\, 2}}{\Box^{\, 2}} \, \prd \prd \cF_{\, 2}\, ,&
&\fr{\pr^{\, 3}}{\Box^{\, 3}} \, \prd \prd \prd \cF_{\, 2}\, ,&
\end{align}
it turns out that
only the first really defines a new structure, because of the Bianchi identity
\be
\prd \cF_{\, 2} \, = \, \fr{1}{4} \, \pr \, \cF_{\, 2}^{\, \pe}\, 
\ee
verified by $\cF_{\, 2}$. 
Thus, we can define for spin $3$ a \emph{one-parameter class} of candidate
Ricci tensors\ft{The subscript ``$\vf$'' in $\cA_{\, \vf}$
($\cA_{\, nl}$ in the notation of \cite{FMS})
is used to distinguish the non-local Ricci tensors from 
their local analogue  to be introduced in Section \ref{section2.1.3}. Those
will be indicated with the symbol
$\cA$, without subscripts, and will depend on the field 
$\vf$ and on  an auxiliary field $\a$.}
\be \label{Ricci3}
\cA_{\, \vf} \, (a_1) \, = \, \cF_{\, 2} \, 
+ \, a_1 \, \fr{\pr^{\, 2}}{\Box} \, \cF_{\, 2}^{\, \pe}\, ,
\ee 
and the very first issue to be clarified is whether the corresponding
postulated equations of motion
\be
\cA_{\, \vf} (a_1) \, = \, 0 \, ,
\ee
can be shown to be consistent, at least for some choices of the parameter $a_1$, 
with the Fronsdal equation $\cF\, = \, 0$, with $\cF$ defined in (\ref{fronsdalT}).
The same issue presents itself in generalised form in the case of spin $s$.
The idea is to consider the ``order zero'' Ricci tensors introduced in 
\cite{fs1, fs2}
 \be \label{kinetic}
   \cF_{n+1} \, = 
     \begin{cases}
    \fr{1}{\Box^{\, n}} \, \cR^{\, [n+1]} \, & \, s \, = \, 2\, (n \, + \, 1) \, , \\
    \fr{1}{\Box^{\, n}} \, \prd \cR^{\, [n]} \, & \, s \, = \, 2\, n \, + \, 1 \, ,
    \end{cases}
 \ee
that might be also defined recursively, according 
to the relation\ft{Here the initial condition is, as already specified, 
$\cF_{\, 1} \, = \, \cF$. It is possible to notice anyway that
the sequence could also formally start with $\cF_{\, 0} \, \eq \, \Box \, \vf$, 
in which case it would produce the Fronsdal tensor $\cF$ as a 
result of the first iteration.}
\be \label{kinetbose}
\cF_{\, n + 1} \, = \, \cF_{\, n} \, + \, \fr{1}{(n + 1)\,(2n + 1)} \, \fr{\pr^{\, 2}}{\Box} \, \cF_{\, n}^{\, \pe} \, 
- \, \fr{\pr}{\Box} \, \prd \cF_{\, n} \, ,
\ee 
and notice that, if the rank of $\vf$ is either $s \, = \, 2\, (n \, + \, 1)$ or 
$s \, = \, 2\, n \, + \, 1$,  they satisfy the series of identities
\be \label{bianchids}
\prd \cF^{\, [k]}_{n+1} \, = \, \fr{1}{2\, (n \, - \, k \, + \, 1)} \, \pr \, \cF^{\, [k + 1]}_{n+1}.
\ee
In particular, in the odd case, $\prd \cF_{n+1}^{\, [n]} \, \eq \, 0$.
These identities imply that divergences of the tensors $\cF_{n+1}$ can always be expressed in terms
of traces, so that the only independent structures that we can consider are 
\be \label{independstruc}
\cF_{n+1} , \ \ \
\cF^{\, \pe}_{n+1}, \ \ \ 
\cF^{\, \pe \pe}_{n+1}, \ \ \
\dots ,\ \ \
\cF^{\, [k]}_{n+1}, \ \ \ 
\dots ,\ \ \
\cF^{\, [q]}_{n+1} \, ,
\ee  
where $q \, = \, n \, + \, 1$ or $q \, = \, n$ depending on whether the rank is
$s \, = \, 2\, (n \, + \, 1)$ or $s \, = \, 2\, n \, + \, 1$.
The most general candidate for a possible Ricci tensor for spin $s$ 
is then given by a linear combination of \emph{all} structures available, 
with coefficients which are arbitrary, up to an overall normalisation

\be \label{genlincomb}
\cA_{\, \vf}\, (\{a_k\}) \, = \, \cF_{n+1} \, + \, 
\dots \, +\, a_k \, \fr{\pr^{\, 2k}}{\Box^{\, k}} \, \cF^{\, [k]}_{n+1} \, + \, \dots \, + \, 
\begin{cases}
 a_{n+1} \, \fr{\pr^{\, 2\, (n+1)}}{\Box^{\, n+1}} \, \cF^{\, [n+1]}_{n+1} & \, s \, = 2\, (n + 1) \, ,\\
 a_{n} \, \fr{\pr^{\, 2\, n}}{\Box^{\, n}} \, \cF^{\, [n]}_{n+1} & \, s \, = 2\, n + 1 \, .
\end{cases}
\ee

The crucial point to be stressed at this level
is that,
\emph{for infinitely many choices of the coefficients $a_1 \, \dots \, a_n$},
the postulated equation $\cA_{\, \vf}\, (\{a_k\}) \, = \, 0$ can be
shown to imply 
an equation of the form
\be \label{compeq1}
\cF \, - \, 3 \, \pr^{\, 3} \, \a_{\, \vf}\, (\{a_k\}) \, = \, 0 \, ,
\ee
where  
$\a_{\, \vf}\, (\{a_k\})$ is a non-local tensor whose 
gauge transformation is a shift in the trace of the 
gauge parameter,
\be \label{shiftalfa}
\d \, \a_{\, \vf}\, (\{a_k\}) \, = \,  \L^{\, \pe} \, ,
\ee
in such a way that, after a suitable gauge-fixing,
infinitely many distinct non-local equations can be reduced to
the Fronsdal\label{notagamma} form\ft{Moreover, as shown in \cite{FMS},
choosing the first coefficient in (\ref{genlincomb})
to be $a_1  = \fr{n}{n + 1}$ the resulting set
of tensors $\cA_{\, \vf}(a_2 \dots a_n)$ can be shown to be already in the 
compensator form (\ref{compeq1}) which,  in this sense, 
it is shown to be highly \emph{not unique}.}. 

 In particular, the reduction to the compensator form (\ref{compeq1})
of the class of equations  $\cF_n \, = \, 0$, 
produces a specific $\a_{\, \vf}$, that we shall denote
$\cH_{\, \vf}$ \cite{fs1, fs2}:
\be \label{oldcompform}
\cF_n \, = \, 0 \ \ \  \Rightarrow \ \ \ \cF \, - \, 3 \, \pr^{\, 3} \, \cH_{\, \vf} \, = \, 0 \, .
\ee
An idea of the mechanism by which (\ref{oldcompform}) is 
realised can be easily obtained for $s = 3$, making the form of  $\cF_2$ explicit, 
\be \label{s3old}
\cF_2 \,=\, \cF \, + \, \frac{1}{6} \, \fr{\pr^{\, 2}}{\Box} \, \cF^{\, \pe} \, - 
\, \fr{1}{2} \, \fr{\pr}{\Box} \, \prd \cF \, = \, 0 \, ,
\ee
and observing that, by virtue of the identity (\ref{fronsdalD}), the
same equation
can be written 
\be \label{220}
\cF_2 \,=\, \cF \, - \, \fr{1}{3} \, \fr{\pr^{\, 2}}{\Box} \, \cF^{\, \pe} \, .
\ee
The trace of  (\ref{s3old}) implies $\cF^{\, \pe} \, = \, \fr{\pr}{\Box} \, \prd \cF^{\, \pe}$
which, upon substitution in (\ref{220}) imply (\ref{oldcompform})
with 
\be
\cH_{\, \vf} \, = \, \fr{1}{3\, \Box^{\, 2}} \, \prd \cF^{\, \pe} \, .
\ee
The general mechanism for (\ref{oldcompform}) is discussed for fermions in Section
\ref{section3.1.3}, and can be adapted to the bosonic case
with minor adjustments\ft{Analogous
results, together with a discussion of
the mixed-symmetric case, were found
in \cite{Bekaert:2002dt, nonsym1, hull, Bekaert:2003zq, bb06}. In particular
in \cite{nonsym1, Bekaert:2003zq} Bekaert and Boulanger, inspired by
previous works \cite{damour, DuboisViolette:1999rd}, showed 
that an equation of the same form of (\ref{oldcompform})
could be deduced from the vanishing of the trace of the curvature,
as a consequence of the generalised Poincar\'e lemma.}.
The main conclusion of this Section is that, starting from the curvatures
(\ref{curvature}), it is possible to define \emph{infinitely many} 
Ricci-like tensors,  according to
(\ref{genlincomb}), all of them unavoidably \emph{non-local}. 
Nonetheless, it is worth stressing that, under the further requirement that 
the kinetic tensors have the lowest possible degree of singularity, the
corresponding ``order-zero'' definition 
(\ref{kinetic}) turns out to be \emph{unique}. This will 
be no more true in general for fermions, as we shall
see in Section \ref{section3.1.2}.
Here we ask ourselves how to obtain a Lagrangian
description for the geometric bosonic theory.

\subsubsection{Geometric Lagrangians} \label{section2.1.3}

 It is a general result that the equation $\cA_{\, \vf}\, (\{a_k\}) \, = \, 0$
is not a Lagrangian equation, but it can
be derived from a Lagrangian \emph{via} a 
multiple-step procedure, once an
identically divergenceless Einstein tensor
is constructed from $\cA_{\, \vf}\, (\{a_k\})$ and its traces.  For the simplest
choice (\ref{kinetic}) this was shown to
be possible  in \cite{fs1, fs2}
where these ``order zero'' Einstein tensors 
were explicitly constructed and look
\be
\cE_{\, \vf}\, = \, \sum_{p \leq n} \, \fr{(-1)^{\, p}}{2^{\, p} \, p\, ! \, 
\left( {n \atop p} \right)} \, \h^{\, p} \, \cF_{\, n}^{\, [p]} \, . 
\ee
Thus, starting with the corresponding Lagrangian
\be
\cL \, = \, \12 \, \vf \, \cE_{\, \vf}\, ,
\ee
it is possible to show that subsequent traces of the Lagrangian equation 
$\cE_{\, \vf}\, = \, 0$ 
imply the condition $\cF_{\, n} \, = \, 0$ and finally, as recalled in the previous Section, 
after some manipulation
involving the identities (\ref{bianchids})
and the gauge-fixing of all non-localities to zero, the Fronsdal equation $\cF = 0$.

 It could be possible to proceed similarly for the generalised Ricci tensors
defined in (\ref{genlincomb}), and in this way infinitely many
non-local, geometric Lagrangians would be 
defined, all describing the same free dynamics. In order to better
understand their meaning, and to look
for a selection principle (if any) inside this class of theories, let us make
some further observations.

 As already stressed, 
the curvatures (\ref{curvature}) define \emph{identically gauge-invariant}
tensors, without the need for algebraic trace conditions
on fields or on gauge parameters. This means that the 
geometric description of higher-spin dynamics is related
to the removal of the constraints
(\ref{fronsdalDT}) and (\ref{fronsdalTG}) assumed in the 
Fronsdal theory, the main drawback being the very 
introduction of non-localities. 

 One different possibility to remove constraints without
introducing non-localities is to replace them, in some sense, with
auxiliary fields. After the first results in this direction \cite{pt}, 
already recalled in the Introduction, 
more recently, a ``minimal'' local formulation
of the same dynamics was proposed in \cite{fs3, fsRev}, whose
building block is in the definition of fully gauge invariant, local, kinetic 
tensors introduced\ft{For the spin-$3$ case an analogous result
had already been found by Schwinger \cite{schwing}.} in
\cite{fs1, fs2, st}. In this setting,
the geometrical meaning of the unconstrained theory, 
obscured by the presence of the auxiliary fields, 
could still be recovered if a clear map between local theory
and non-local ones could be established. This was done in 
\cite{FMS}, and it is briefly reviewed in the following,
with focus on the bosonic case.

For a rank-$s$ fully symmetric tensor one can begin by
considering the Fronsdal tensor $\cF$ introduced in (\ref{fronsdalT})
and compensating its gauge transformation
\be \label{transfgauge}
\d \, \cF \, = \, 3 \, \pr^{\, 3} \, \L^{\, \pe} \, ,
\ee
with the introduction of a spin-$(s-3)$ \emph{compensator} $\a$ transforming as
\be \label{transalfa}
\delta \, \a \, = \, \Lambda^{\, \prime}\, ,
\ee
so that the local kinetic tensor\ft{As previously advertised, 
we distinguish the local tensor $\cA$, function of $\vf$ and $\a$, 
without subscripts, from 
the non-local tensors $\cA_{\, \vf} (\{a_k\})$ defined in
(\ref{genlincomb}) , function of $\vf$ only.
Similarly, $\a$ without subscript is an independent field, 
whereas $\a_{\, \vf}$ indicates in general a non-local tensor, function
of $\vf$, with the property (\ref{shiftalfa}).}
\be \label{tensorA}
\cA \, = \, {\cF} \, - \, 3 \, \pr^{\, 3} \, \alpha \, ,
\ee
be \emph{identically gauge-invariant}.
The Bianchi identity
\be \prd \cA \, - \, \frac{1}{2} \, \pr \, \cA^{\, \prime} \, = \, -
\, \frac{3}{2} \, \pr^{\, 3} \, \left(\vf^{\, \prime \prime} \, - \,
4 \, \pr \cdot \a \, - \, \pr \, \a^{\, \prime} \right)\ ,
\label{bianchibose} \ee
is the other main ingredient needed to show that a gauge-invariant 
local Lagrangian can be written in the compact form
\be \label{boselagr} 
\cL \, = \, \frac{1}{2} \, \vf \, \left(\cA
\, - \, \frac{1}{2} \, \h \, \cA^{\, \pe} \right) \, - \,
\frac{3}{4} \ {s \choose 3 } \, \a\, \prd \cA^{\, \pe} \, + \, 3
\, { s \choose 4 } \, \beta \, \left[ \vf^{\, \pe \pe} \, - \,
4 \, \prd \a \, - \, \pr \, \a^{\, \pe} \right]\, , 
\ee
where the \emph{Lagrange multiplier} $\beta$
transforms 
as\ft{It is possible to show that $\b$ is necessary in 
order for the double trace not to propagate
\cite{FMS}, so that
the field content represented by the triple 
$\vf, \a, \b$ is the minimal one needed to remove
the constraints from the Fronsdal Lagrangian \cite{fs3}.}
\be \label{transbeta} 
\delta \beta \, = \, \prd \prd \prd \Lambda \, ,
\ee
and the tensor $\cC \, \eq \, \vf^{\, \pe \pe} \, - \,
4 \, \prd \a \, - \, \pr \, \a^{\, \pe}$ is identically
gauge-invariant.

 In order to review in which sense a link between 
the local theory described by (\ref{boselagr}) and the 
geometric Lagrangians can be established, it will 
be sufficient to analyse the case of spin $4$. 
A more complete discussion can be found in \cite{FMS}.

 In the spin-$4$ case, the Lagrangian 
equations\ft{The ``compensator'' tensor $\cA = \cF - 3 \, \pr^{\, 3} \, \a$ also plays a role in the linearised 
Vasiliev's equations, if a suitable, unusual, projection is performed
\cite{sss}. More recently, in a novel approach
to the quest for a higher-spin action principle, 
proposed in terms of a Chern-Simons theory, it has been 
shown that the equation $\cA = 0$ is the natural outcome, at least
for the spin-$3$ case,
of the linearisation procedure in  that dynamical framework
\cite{joan, joan2}.}
coming from (\ref{boselagr}) can be simplified to
\be \label{leqspin4}
\begin{split}
&\cA\, - \, \12 \, \h\, \cA^{\, \pe} \, = \, 0     \\
&\prd \cA^{\, \pe} \, = \, 0 \, \\
& \vf^{\, \pe \pe} \, = \, 4 \, \prd \, \a \, .
\end{split}
\ee
We can invert the second equation to find $\tilde{\a}_{\, \vf}$ as the non-local
solution for the compensator in terms of the basic field; the result is
\be \label{comp4}
\tilde{\a}_{\, \vf} \, = \, \fr{1}{3 \, \Box^{\, 2}} \, \prd \cF^{\, \pe} \, - \, 
\fr{\pr}{4 \, \Box^{\, 3}} \, \prd \prd \cF^{\, \pe} \, ,
\ee
which, upon substitution in (\ref{leqspin4}) implies the non-local \emph{system}
\be \label{nleqspin4}
\begin{split}
&\tilde{\cA}_{\, \vf} \, - \, \12 \, \h\, \tilde{\cA}^{\, \pe}_{\, \vf} \, = \, 0 \, , \\
& \vf^{\, \pe \pe} \, = \, 4 \, \prd \, \tilde{\a}_{\, \vf} \, ,
\end{split}
\ee
where  $\tilde{\cA}_{\, \vf}  =  \cF -  3 \, \pr^{\, 3} \, \tilde{\a}_{\, \vf} $.
From the discussion of the previous Section 
we also know that, starting with the geometric theory defined in terms
of the simplest Ricci tensor (\ref{kinetic3,4}), the result is an equation of motion
of the form $\cF_2 \, = \, 0$,
that can be shown to imply the compensator-like equation
\be \label{comp4h}
\cF \, - \, 3 \, \pr^{\, 3} \, \cH_{\, \vf} \, = \, 0 \, ,
\ee
with $\cH_{\, \vf}$ \emph{different} from $\tilde{\a}_{\, \vf}$ in 
(\ref{comp4}), and given  by
\be
\cH_{\, \vf} \, = \, \fr{1}{3 \, \Box} \, \prd \cF^{\, \pe} \, - \, \fr{1}{4} \, \fr{\pr}{\Box^{\, 2}}
\cF^{\, \pe \pe} \,  .
\ee
The very fact  that $\tilde{\a}_{\, \vf}$ and $\cH_{\, \vf}$
do not coincide,
their difference being a gauge invariant tensor,
is another way to see the infinite degeneracy of the
free geometric theory.
Actually, both the first equation in (\ref{nleqspin4}) 
and (\ref{comp4h})
can be made Lagrangian by the construction of suitable Einstein tensors, 
say $\cE_{\, \tilde{\a}}$ and $\cE_{\, \cH}$,
but then (almost) any linear combination  of these two Einstein tensors 
with coefficients $a$ and $b$ such that 
$a + b = 1$, is again 
an allowed tensor for a non-local Lagrangian whose dynamics will be equivalent 
to the Fronsdal one. 

 From the viewpoint of the connection we are after, 
between local and non-local theories, it is important to 
stress that a proper counterpart of the
local theory should involve a Ricci tensor 
$\cA_{\, \vf}$ having the properties encoded in \emph{the system} 
(\ref{nleqspin4}), and that
neither $\tilde{\cA}_{\, \vf}$ alone nor $\cF -  3  \, \pr^{\, 3} \, \cH_{\, \vf}$ 
satisfy this requirement. For instance, 
whereas  ${\tilde{\cA}}^{\, \pe \pe}_{\, \vf}$ and  
$(\cF -  3  \, \pr^{\, 3} \, \cH_{\, \vf})^{\, \pe \pe}$ 
do not vanish, it is possible to show that the 
second equation in (\ref{nleqspin4}) effectively implies that the tensor 
$\tilde{\cA}_{\, \vf}$ is \emph{identically doubly traceless}
regardless the fact that the first equation be satisfied or not, i.e. even if $\vf$ is 
off-shell.

 One possibility to get closer to (\ref{nleqspin4})
is then to select the particular combination of $\cE_{\, \tilde{\a}}$ and $\cE_{\, \cH}$
such that in the resulting Einstein tensor the non-local ``compensator block''
\be \label{G}
\cA_{\, \vf} \, \eq \, \cF \, - \, 3 \, \pr^{\, 3} \, \g_{\, \vf} \, 
\ee
be \emph{identically} doubly-traceless. 
The unique solution gives in this particular case an Einstein 
tensor of the form
\be
\cE_{\, \vf} \, = \, \fr{4}{3} \, \cE_{\, \tilde{\a}} \, - \, \fr{1}{3} \,  \cE_{\, \cH} \, = \, 
\cA_{\, \vf} \, - \, \12 \, \h \, \cA_{\, \vf}^{\, \pe} \, + \, \h^{\, 2} \, 
\cB_{\, \vf} \, ,
\ee
with $\cB_{\, \vf}$ such that
\be
\pr \, \cB_{\, \vf} \, = \, \12 \, \prd \cA_{\, \vf}^{\, \pe} \, .
\ee
and $\g_{\, \vf}$ given by
\be
\g_{\, \vf} \, = \, \fr{1}{3\, \Box^{\, 2}} \prd \cF^{\, \pe} \, - \, 
\fr{1}{3}\, \fr{\pr}{\Box^{\, 3}} \prd \prd \cF^{\, \pe} \, + \, 
\fr{1}{12}\, \fr{\pr}{\Box^{\, 2}}\,\cF^{\, \pe \pe} \, .
\ee

 This analysis of the spin-$4$ case suggests that, among the 
infinitely many geometric theories dynamically equivalent
to the Fronsdal constrained system, \emph{only one}
should be identified as the proper non-local counterpart
of the local theory defined by the Lagrangian (\ref{boselagr}).

 A crucial observation in order to corroborate this hypothesis,
and to generalise the result to all spins,
is related to the analysis of the current exchange
in the presence of weak external sources performed in 
\cite{FMS}. There it was shown that the correct
structure of the propagator is guaranteed if and only if
the Einstein tensor has the form, for any spin $s$,
\be \label{einstein}
\cE_{\, \vf} \, = \, \cA_{\, \vf} \, - \, \12 \, \h \, \cA_{\, \vf}^{\, \pe} \, + \, \h^{\, 2} \, 
\cB_{\, \vf} \, ,
\ee
with $\cA_{\, \vf}$ given by (\ref{G}), and satisfying the two identities
\be \label{Aident}
\begin{split}
& \prd \cA_{\, \vf} \, - \, \12 \, \pr \, \cA_{\, \vf}^{\, \pe}  \, \eq \, 0 \, , \\
& \cA_{\, \vf}^{\, \pe \pe}  \, \eq \, 0 \, ,
\end{split}
\ee
whereas the general requirement that the Einstein
tensor be divergenceless fixes the tensor $\cB_{\, \vf}$ in terms
of $\cA_{\, \vf}$. 
The explicit dependence of $\cA_{\, \vf}$ on the curvatures
has been given 
in \cite{FMS} and looks\ft{We correct here a misprint in the 
corresponding equation ($4.67$) in \cite{FMS}.}
\be \label{geomA}
\cA_{\, \vf} \, = \, \cF \, -\, 3 \, \pr^{\, 3} \, \g_{\, \vf} \, = \, 
\sum_{k=0}^{n+1} \, (-1)^{k+1} \, (2\, k\,  -\, 1)\, 
\{ \fr{n\, + \, 2}{n\, - \, 1} \, \prod_{j=-1}^{k-1}\, \fr{n \, + \, j}{n \, - \, j \, + \, 1} \} \, \fr{\pr^{\, 2k}}{\Box^{\, k}}\,
\cF_{n+1}^{\, [k]} \, ,
\ee
with $\cF_{\, n+1}$ defined in (\ref{kinetic}).
Details about the dependence of $\cA_{\, \vf}$ and $\cB_{\, \vf}$ 
on the Fronsdal tensor $\cF$ are
given in Section $4$ and in Appendix \ref{B}, when discussing the 
relation between this geometric solution and the  
generalised Fierz-Pauli mass terms introduced in Section \ref{section2.2.3}.
 
 Here we further observe that this result is particularly meaningful, 
in that not only it implies  the existence of a clear, 
one-to-one map between minimal local theory
and \emph{one} geometric formulation, but it also 
gives a \emph{physical meaning} to this map, 
in terms of consistency of the coupling with external sources.

 Nonetheless, the local counterpart of the geometric theory,
defined by (\ref{boselagr}), contains higher derivatives, in the kinetic
operator of the compensator field $\a$, which could be 
seen as another facet of the difficulties met 
in a geometric-inspired description of the dynamics. 

On the other hand, the very fact that the field $\a$ can be 
removed from the equations of motion by going to the 
``Fronsdal'' gauge, where the parameter $\L$ is traceless,
indicates that the physical content of the theory
should be safe from difficulties related to the 
higher-derivative terms.

 Thus, to give further support to this viewpoint,
before discussing the issue of constructing suitable mass
deformations for the geometric theory, we shall
show how the local counterpart of the geometric
description can be put in more conventional form, 
constructing an equivalent, but ordinary-derivative, Lagrangian.

\subsubsection{Ordinary-derivative Lagrangians for unconstrained bosons} \label{section2.1.4}

 We would like to investigate the possibility of eliminating the higher-derivative
terms in the minimal Lagrangians, while still retaining their dynamical content.

  The basic idea is to look for ``compensators'' transforming as 
\emph{gradients} of the trace of the gauge 
parameter\ft{I am grateful to J. Mourad for discussions about this point.}. 
This choice does not lead to a straightforward
solution, to begin with just because the new compensators 
are no more pure-gauge fields, and it is not obvious how to avoid 
their propagation. 

 The solution  is simply to include a constraint in the Lagrangian
so as to ``remember'' that the lower-derivative compensator
actually ``is'' a gradient of $\a$, whereas some more 
attention has to be paid to the role of the Lagrange
multipliers of the theory, in order to make sure
that they do not propagate extra degrees of freedom.

 The starting point is the gauge transformation of the Fronsdal
tensor:
\be
\d \, \cF \, = \, 3 \, \pr^{\, 3} \, \L^{\, \pe} \, .
\ee 
In order to define an unconstrained, local kinetic tensor, instead
of the field $\a$, let us consider the alternative possibility
\be
\th : \ \ \ \d \, \th \, = \, \pr \,  \L^{\, \pe} \, ,
\ee
and define the corresponding gauge-invariant tensor 
according to\ft{Here we use a double subscript, to avoid
possible confusion with the local tensor $\cA = \cF - 3 \, \pr^{\, 3} \, \a$, 
and with the non-local tensor, function of $\vf$ alone, 
$\cA_{\, \vf} = \cF - 3 \, \pr^{\, 3} \, \g_{\, \vf}$
given by (\ref{geomA}).}
\be
\cA_{\, \vf, \,\th} \, = \, \cF \, - \, \pr^{\, 2} \, \th \, .
\ee
Clearly, this choice runs into the trouble 
of introducing in the theory a field which is not a pure shift. 
This problem can be solved by simply including in the
Lagrangian   a suitable
constraint relating the field $\th$ and the compensator $\a$.
We can start with a trial Lagrangian of the  form, 
\be \label{trial}
\cL_0 \, = \, \12 \, \vf \, \{\cA_{\, \vf, \,\th} \, - \, \12 \, \h \, \cA_{\, \vf, \,\th}^{\, \pe}\}
\, + \, {s \choose 2} \, \g \, (\th \, - \, \pr \, \a) \, ,
\ee
with $\g$ a gauge-invariant Lagrange multiplier, whose
normalisation has been chosen 
for future purposes.

 We can already notice the (obvious) point which will play a 
crucial role in what follows: since $\a$ is anyway present in 
$\cL$, we are free to use it in other combinations, if needed. 

 The second structure we need is the Bianchi identity for $\cA_{\, \th}$
\be
\prd \cA_{\, \vf, \,\th} \, - \, \12 \, \pr \, \cA_{\, \vf, \,\th}^{\, \pe} \, 
= \, - \, \fr{1}{2}\, \pr \, \{\pr^{\, 2} \, \vf^{\, \pe \pe} \, + \, \Box \, \th \, - \, 
\pr \, \prd \th \, -  \, \pr^{\, 2} \, \th^{\, \pe}\}\, \eq \, - \12 \, \pr \, \hat{\cC} \, ,
\ee
from which it is already possible to observe the second 
(and more delicate) difficulty of this approach: 
the structure of the gauge-invariant
combination of fields to be compensated in the 
variation of $\cL_0$
involves a $\Box$ of the field $\th$. This implies that
the corresponding Lagrange multiplier that one would 
introduce by analogy with (\ref{boselagr}) would appear as a propagating
field in the equations of motion for the $\th$ itself.

 More explicitly, let us compute the variation of 
the trial Lagrangian (\ref{trial}), 
\be
\d \, \cL_0 \, = \, - \, \fr{1}{4} \, {s \choose 2} \, \pr \, \L^{\, \pe}
\, \cA_{\, \vf, \,\th}^{\, \pe} \, - \, \12 \, {s \choose 2} \, \prd  \L \, \hat{\cC}\, ,
\ee
and to begin with, in order to make the problem
related with this choice explicit, let us try to compensate the $\hat{\cC}$-term 
introducing  a multiplier 
\be
\hat{\b} :\ \ \  \d \, \hat{\b} \, = \, \prd \L \, ,
\ee
allowing to complete the construction of a gauge invariant Lagrangian  
according to
\be
\cL \, = \, \12 \, \vf \, \{\cA_{\, \vf, \,\th} \, - \, \12 \, \h \, \cA_{\, \vf, \,\th}^{\, \pe}\}
\, + \, \fr{1}{4} \, {s \choose 2} \, \th \, \cA_{\, \vf, \,\th}^{\, \pe} \, 
+ \, \12 \, {s \choose 2} \, \hat{\b} \, \hat{\cC} \, + \, 
{s \choose 2} \, \g \, (\th \, - \, \pr \, \a) \, .
\ee
The corresponding  equations of motion are
\begin{align}
& E_{\, \vf}\, : & & \ \  \cA_{\, \vf, \,\th} \, - \, \12 \, \h \, (\cA_{\, \vf, \,\th}^{\, \pe}\,
- \, \12 \, \hat{\cC}) \, + \, \h^{\, 2} \, \hat{\cB} \, = \, 0 \, , & \nonumber \\
& E_{\, \th}\, : & & \ \ 2 \, \cA_{\, \vf, \,\th}^{\, \pe} \, 
- \, \hat{\cC} \, - \, \Box \, \hat{\cD} \, + \, \pr \, \prd \hat{\cD} \, - \, 
2 \, \h \, \hat{\cB} \, + \, 4 \, \g \, = \, 0 \, ,& \nonumber \\
&E_{\, \a}\, : & & \ \ \, \prd \g \, = \, 0 \, , \nonumber \\
& E_{\, \b_1}\, :& & \ \  \, \hat{\cC} \, = \, 0 \, , &\nonumber \\
& E_{\, \g} \, : & & \ \ \th \, - \, \pr \, \a \, = \, 0 \, ,& 
\end{align}
where the various tensors are defined by
\be
\begin{split}
& \cA_{\, \vf, \,\th} \, = \, \cF \, - \, \pr^{\, 2} \, \th \, , \\
& \hat{\cB} \, = \, \prd \prd \hat{\b} \, - \, \12 \, (\prd \prd \vf^{\, \pe} \, - \, \prd \prd \th) \, , \\
&\hat{\cC} \, = \, \pr^{\, 2} \, \vf^{\, \pe \pe} \, + \, \Box \, \th \, 
- \, \pr \, \prd \, \th \, - \, \pr^{\, 2} \, \th^{\, \pe} \, , \\
&\hat{\cD} \, = \, \vf^{\, \pe} \, - \, \th \, - \, 2 \, \hat{\b} \, ,
\end{split}
\ee
and the gauge transformations of the fields are
\be
\begin{split}
&\d \, \vf \, = \, \pr \, \L \, , \\
&\d \, \th \, = \, \pr \, \L^{\, \pe} \, , \\
&\d \, \hat{\b} \, = \, \prd \L \, , \\
&\d \, \a \, = \, \L^{\, \pe} \, , \\
&\d \, \g \, = \, 0 \, .
\end{split}
\ee
Now, given that in the gauge $\a = 0$ the equation for $\vf$ is manifestly 
consistent\ft{In this gauge $\cA_{\, \vf, \,\th}^{\, \pe \pe} = 0$, and the 
double trace of  $E_{\, \vf}$, implying $\hat{\cB} = 0$, simply
fixes the double divergence of $\hat{\b}$ in terms of $\vf$.}
on the other hand it is difficult to avoid the conclusion that $\hat{\b}$ is 
a propagating field as well, given that it is not possible to express it completely
in terms of the other fields, and because of the presence of $\Box \, \hat{\cD}$ in 
$E_{\, \th}$\ft{From this point of view, the cancellation of the terms
in $\Box \, \a$ in the Bianchi identity of the minimal theory
(\ref{boselagr}), 
that would have led to the same problem, without possible solutions, looks
somewhat magical. Of course, the ``magic'' is in the quasi-conservation
of the Einstein tensor in the constrained setting, implying that 
only gradients of $\vf^{\, \pe \pe}$ can appear, and then
the structure in $\a$ follows from the gauge transformation of 
$\vf^{\, \pe \pe}$.}.

 Nonetheless, we can think of a possible way out, 
substituting $\hat{\b}$ with the following combination\ft{And \emph{not} 
$\hat{\b} \  \ra \  \12\, (\vf^{\, \pe} \, - \, \pr\, \a)$, that would 
give a higher-derivative term in $\cL$.} 
\be
\hat{\b} \, \ \ \ra \ \ \, \12\, (\vf^{\, \pe} \, - \, \th) \, .
\ee
together with the addition of the further coupling of the form
$\sim \b \, \{\vf^{\, \pe \pe} - 4 \, \prd \a - \pr \, \a^{\, \pe}\}$, 
meant to provide the correct meaning of the 
double-trace of $\vf$, that would be no more under control
in the absence of a true constraint equation.

 The complete Lagrangian is
\be
\begin{split} \label{basiclagr}
\cL \, = & \, \12 \, \vf \, \{\cA_{\, \vf, \,\th} \, - \, \12 \, \h \, \cA_{\, \vf, \,\th}^{\, \pe}\}
\, + \, \fr{1}{4} \, {s \choose 2} \, \th \, \cA_{\, \vf, \,\th}^{\, \pe} \, 
+ \, \fr{1}{4} \, {s \choose 2} \,  (\vf^{\, \pe} \, - \, \th) \, \hat{\cC} \, \\
&  + \, {s \choose 2} \, \g \, (\th \, - \, \pr \, \a) \, + \, 
3 \, {s \choose 4} \,\b \,  \{\vf^{\, \pe \pe} \, - \, 4 \, \prd \, \a \, - \, \pr \, \a^{\,\pe}\} \, ,
\end{split}
\ee
with $\b$ a \emph{gauge-invariant} multiplier. The corresponding equations
are
\begin{align} \label{basiceom}
& E_{\, \vf}\, : & & \ \  \cA_{\, \vf, \,\th} \, - \, \12 \, \h \, (\cA_{\, \vf, \,\th}^{\, \pe}\,
- \, \hat{\cC}) \, + \, \h^{\, 2} \, \b \, = \, 0 \, , & \nonumber \\
& E_{\, \th}\, : & & \ \ \cA_{\, \vf, \,\th}^{\, \pe} \, 
- \, \hat{\cC} \, + \, 2 \, \g \, = \, 0 \, ,& \nonumber \\
&E_{\, \a}\, : & & \ \ \prd \g \, 
+ \, \pr \, \b \, + \, \12 \, \h \, \prd \b \, = \, 0 \, ,  \\
& E_{\, \b}\, :& & \ \  \vf^{\, \pe \pe} - 4 \, \prd \a - \pr \, \a^{\, \pe}\, = \, 0 \, , &\nonumber \\
& E_{\, \g} \, : & & \ \ \th \, - \, \pr \, \a \, = \, 0 \, ,\nonumber& 
\end{align}
and the set of gauge transformations is given by
\be
\begin{split}
&\d \, \vf \, = \, \pr \, \L \, , \\
&\d \, \th \, = \, \pr \, \L^{\, \pe} \, , \\
&\d \, \a \, = \, \L^{\, \pe} \, , \\
&\d \, \b \, = \, 0 \, , \\
&\d \, \g \, = \, 0 \, .
\end{split}
\ee
Let us make a few comments:
\begin{description}
 \item[$\ra$] the 
equation for $\g$ transforms $\cA_{\, \vf, \,\th}$ and $\hat{\cC}$ in the corresponding
quantities of (\ref{boselagr}):
 \be
 E_{\, \g} \, \Rightarrow
 \begin{cases}
  \cA_{\, \vf, \,\th} \, \ra \, \cF \, - \, 3 \, \pr^{\, 3} \, \a \, ,\\
  \hat{\cC} \, \ra \, \vf^{\, \pe \pe} - 4 \, \prd \a - \pr \, \a^{\, \pe} \, ,
 \end{cases}\ee
this in its turn implies $ E_{\, \vf}\ \ \ra \ \  \cA \, - \, \12 \, \h \, \cA^{\, \pe}
\, + \, \h^{\, 2} \, \b \, = \, 0$, which is in fact the Lagrangian equation of (\ref{boselagr}), 
with some specification to be given about the multiplier $\b$. 
 \item[$\ra$] $\b$ is a gauge-invariant tensor\ft{Playing somehow the role
 of the tensor $\cB$ of the minimal theory of
 \cite{FMS}.}. Since $\cC = 0 \ \ra \ \cA^{\, \pe \pe} = 0$, it is simple
to realise that multiple traces of $E_{\, \vf}$ imply that all traces of $\b$, and finally
$\b$ itself, vanish in the free case. In the presence of a current $\cJ$,
$\b$ would be fixed in terms of $\cJ^{\, \pe \pe}$ .
 \item[$\ra$] $\g$ is determined in terms of the other fields, and in particular
 in the gauge $\a = 0$ it is proportional to $\cF^{\, \pe}$.
 \item[$\ra$ \, ] Consistency with gauge-invariance can be expressed by the 
 identity
 \be
 \prd E_{\, \vf} \, = \, \h \, \{\fr{1}{3 \, {s \choose 3}} \, E_{\, \a} \, -
 \fr{1}{{s \choose 2}} \, \prd E_{\, \th} \} ,
 \ee
where prefactors coming from the variation of (\ref{basiclagr}), 
neglected in (\ref{basiceom}), have also been taken into account.
\end{description} 

 In this sense, at the price of 
enlarging the field content of the minimal theory to include the new fields 
$\th$ and $\g$ we can characterise 
the same dynamics as (\ref{boselagr}) 
by means of the ordinary-derivative Lagrangian (\ref{basiclagr}). 
As already recalled in the Introduction, the difference
with respect to the recent result found in 
\cite{Buchbinder:2007ak}, similar in spirit and in the
total number of fields involved, is that  the
Lagrangian (\ref{basiclagr}) somehow represents
the ordinary-derivative version of the geometric
theory synthetically described by (\ref{einstein}), (\ref{Aident}) and 
(\ref{geomA})\ft{It is conceivable, and it represents
an interesting issue to be clarified, that the Lagrangians (\ref{boselagr}) and 
(\ref{basiclagr}), together with the corresponding one introduced in 
\cite{Buchbinder:2007ak}, all encoding the same 
irreducible dynamics, might be related
by some kind of field redefinition. On the other hand, 
the very fact that the field content is similar, but
not identical (the Lagrangian of \cite{Buchbinder:2007ak}
involving a total of six fields), makes it not directly obvious
which could be the possible redefinition allowing to switch among
these possibilities, \emph{off-shell}. I would like 
to thank the Referee for stimulating this comment.}.

 In the next Section we turn our attention again to the geometry, 
and start to investigate the possibility of using the geometric Lagrangians
for the study of the massive representations.

\subsection{Mass deformation} \label{section2.2}

  We look for a massive Lagrangian
for higher-spin bosons of the form
\be \label{geomlagmass}
\cL \, = \,  \12 \, \vf \, \{\cE_{\, \vf} - \, m^{\, 2} \, M_{\, \vf}\}  \, ,
\ee
with $\cE_{\vf}$ a \emph{generic} member in the 
class of divergence-free
Einstein tensors discussed in Section \ref{section2.1.3}, and $M_{\, \vf}$  a linear
function of $\vf$ to be determined.

 The main idea is that $M_{\, \vf}$ should be a linear
combination of \emph{all} the traces of $\vf$, 
starting with the  Fierz-Pauli mass term (\ref{FPspin2}), 
that we rewrite here for convenience
\be \label{fpmass}
M_{\, FP} \, = \, \vf \, - \, \h \, \vf^{\, \pe} \, .
\ee
Qualitatively speaking this is plausible, 
in the sense that there is no reason in principle to assume that 
only order-zero and order-one traces should contribute in the 
unconstrained case.
In general, however, the coefficients of the various terms
in the sequence could be spin-dependent, so that,  
for instance in the spin-$3$ case, the mass term could take the form
\be
M_{\, s = 3} \, = \, \vf \, - \, k \, \h \, \vf^{\, \pe} \, ,
\ee
with a given constant $k$.
For  the \emph{constrained} case it was shown in \cite{ADY87} that
$k = 1$ is the only acceptable value for \emph{all} spins. 
This is particularly clear if one considers the dimensional reduction
of the constrained massless theory from $D + 1$ to $D$ dimensions.
In that framework indeed the very form of the Fronsdal Lagrangian
\be
\cL \, = \, \12 \, \vf \, \{\cF \, - \, \12 \, \h \, \cF^{\, \pe}\} \,  \sim \, 
\12 \, \vf \, \{\Box \, (\vf \, - \, \h \, \vf^{\, \pe}) \, + \, \dots\} 
\ee
implies that, under the formal substitution
\be
\Box \, \ra \, \Box \, - \, m^{\, 2} \, ,
\ee
the mass term will appear exactly in the Fierz-Pauli form, for all spins.
The same result can be found in the Kaluza-Klein reduction of the
unconstrained, local theory of 
\cite{fs3, fsRev, FMS}, with a richer structure
of Stueckelberg fields, as expected in order to account for the wider gauge symmetry
allowed in that context.

 The passage from the local description to the non-local one 
can be roughly described by the requirements that the compensator
$\a$ be replaced by a suitable non-local tensor with the same gauge 
transformation, and that higher-traces of the field enter the 
Lagrangian, to replace the equation of motion for the 
multiplier $\b$, ensuring in the local setting that $\vf^{\, \pe\pe}$
be pure gauge. No modifications are expected for the 
local, lower-trace parts of the theory, and in this sense
we do not expect the Fierz-Pauli term (\ref{fpmass}) to
be modified, if not for the  contribution of 
further traces of the field. 

 More quantitatively, we shall see that, starting from the equation 
\be
\cE_{\vf}  \, - \, m^{\, 2} \, M_{\, \vf} \, = \, 0 \, ,
\ee
a necessary
condition in order to recover the Fierz system (\ref{fierz})
will be that 
$\cA^{\, \pe}_{\, \vf}\, (\{a_k\})$, as can be computed
from (\ref{genlincomb}),  vanish on-shell,
and to this end we shall need \emph{exactly}
the Fierz-Pauli constraint
\be \label{fpagain}
\prd \vf \, - \, \pr \, \vf^{\, \pe} \, = \, 0 \, ,
\ee
whereas any other deformation of that condition, of the type
\be \label{fpdef}
\prd \vf \, - \, k \, \pr \, \vf^{\, \pe} \, = \, 0 \, ,
\ee
that would come from a different form of the mass term, with $k \neq 1$, would not work. 

 Once it is recognised that the crucial condition to reach is
(\ref{fpagain}), the whole remainder of the sequence in $M_{\, \vf}$
has to be fixed in such a way that the equation
\be
\prd M_{\, \vf} \, = \, 0 \, 
\ee
yield (\ref{fpagain}), \emph{together with all the
consistency conditions} coming from the traces of (\ref{fpagain}) itself.

 We begin by displaying the strategy in the simpler cases 
of spin $3$ and spin $4$, to move then in Section \ref{section2.2.3} to the 
general case.  

 \subsubsection{Spin $3$}

 Since for spin $3$ there are no further traces after the first, 
we assume the Fierz-Pauli mass term, and consider the massive Lagrangian
\be \label{lagrs3}
\cL \, = \, \12 \, \vf \, \{\cE_{\vf}  \, -  \, m^{\, 2} \, 
(\vf \, - \, \h \, \vf^{\, \pe})\} \, .
\ee
Taking a divergence, and thereafter a trace of the corresponding equation of motion
\be 
\cE_{\vf} \, -  \, m^{\, 2} \, 
(\vf \, - \, \h \, \vf^{\, \pe}) \, = \, 0 \, ,
\ee
we obtain first
\be
\prd \vf^{\, \pe} \, = \, 0 \, ,
\ee
and then, as desired, (\ref{fpagain}).
These two conditions imply, in this case, that the trace of the
Fronsdal tensor $\cF$
\be
\cF^{\, \pe} \, = \, 2 \, \Box \, \vf^{\, \pe} \, - \, 2 \, \prd \prd \vf \, + \, \pr \,
\prd \vf^{\, \pe} \, ,
\ee
vanishes on-shell, together with the trace of the 
``elementary'' Ricci tensor (\ref {kinetic3,4}),
$\cF_2^{\, \pe}$, 
that always contains at least one trace of $\cF$, 
as it is obvious from (\ref{220}). 

 Given the form (\ref{Ricci3}) of the 
general candidate Ricci tensor for spin $3$, that we
report here for simplicity
\be 
\cA_{\, \vf} \, (a_1) \, = \, \cF_{\, 2} \, 
+ \, a_1 \, \fr{\pr^{\, 2}}{\Box} \, \cF_{\, 2}^{\, \pe}\, ,
\ee
the conclusion is that \emph{whatever geometric Einstein
tensor we choose in} (\ref{lagrs3}), after the implementation
ot the Fierz-Pauli constraint the resulting
equation  will anyway be 
\be
\cF_2 \, - \, m^{\, 2} \, (\vf \, - \, \h \, \vf^{\, \pe}) \, = \, 0 \, ,
\ee
whose trace implies $\vf^{\, \pe} = 0$, then $\prd \vf = 0$, and finally the 
Klein-Gordon equation, given that under these conditions
the full Ricci tensor $\cF_2$ reduces to $\Box \, \vf$.
It is already possible to appreciate the special role played by (\ref{fpagain}):
in order to make $\cF^{\, \pe}$  vanish, we \emph{need} the
condition 
\be
\Box \, \vf^{\, \pe} \, - \,  \, \prd \prd \vf \, = \, 0 \, ,
\ee
which can only be a consequence of (\ref{fpagain}), and cannot 
be derived from any other different relation
of proportionality between $\prd \vf$ and $\pr \, \vf^{\, \pe}$.

 \subsubsection{Spin $4$}

For spin $s \geq 4$, as previously discussed, we consider reasonable to 
include in the mass term further traces of the field.
We assume then, for $s = 4$, the Lagrangian
\be
\cL \, = \, \12 \, \vf \, \{\cE_{\vf}  \, 
- \, m^{\, 2} \, M_{\, \vf} \}\, ,
\ee 
where, in general,
\be
M_{\, \vf} \, = \, \vf \, + \, a \, \h \, \vf^{\, \pe} \, + b \, \h^{\, 2} \, \vf^{\, \pe \pe}\, .
\ee
Again, we would like to fix the coefficients in the mass term
so that, on-shell, 
\be
\cA_{\vf}^{\, \pe}\, (a_1, a_2) = (1\, + \, a_1) \, \cF_2^{\, \pe}\, + \,  
(3\, a_1 \,+ \, a_2)\, \fr{\pr^{\, 2}}{\Box} \, \cF_2^{\, \pe \pe} \, = \, 0 \, ,
\ee
at least for some choices of $a_1$ and $a_2$.
On the other hand, from
the explicit form of $\cF_2$
\be 
\cF_2 \,=\, \cF \, - \, \frac{1}{3} \, \fr{\pr^{\, 2}}{\Box} \, \cF^{\, \pe} \, + 
\, \fr{\pr^{\, 4}}{\Box^{\, 2}} \, \cF^{\, \pe \pe} \, ,
\ee
it is possible to see that $\cA_{\vf}^{\, \pe}$ starts with $\cF^{\, \pe}$ together with
terms containing at least one divergence of $\cF^{\, \pe}$.
As a consequence of this fact, in $\cF^{\, \pe}$
\be \label{Fprime}
\cF^{\, \pe} \, = \, 2 \, \Box \, \vf^{\, \pe} \, - \, 2 \, \prd \prd \vf \, + \, \pr \,
\prd \vf^{\, \pe} \, + \, \pr^{\, 2} \, \vf^{\, \pe \pe} \, ,
\ee
the first two terms cannot be compensated by anything in
the remainder of $\cA_{\vf}^{\, \pe}$.
This means that, in order for the program to be realised, the combination
\be
\Box \, \vf^{\, \pe} \, -  \, \prd \prd \vf \, 
\ee
has to be expressible in terms
of higher traces and divergences of $\vf$,  
as a consequence of the equations of motion.
This kind of condition, in turn, is implemented by the Fierz-Pauli constraint, 
and would not hold if the constraint had the more general form (\ref{fpdef})
with $k \neq 1$.

 If we then assume to have fixed the coefficients 
$a$ and $b$ so that $\prd M_{\, \vf} \, = \, 0$ implies (\ref{fpagain}), the 
following consequences can be shown to hold:
\begin{align} %\label{onshellid}
& \cF \, = \, \Box \, \vf \, - \, \pr^{\, 2} \vf^{\, \pe} \, , &&&&\cF^{\, \pe} \, = 3 \, \pr^{\, 2} \, \vf^{\, \pe \pe}\, ,& \nonumber\\
&\cF_2 \, = \, \cF \, -\, 3\, \pr^{\, 4} \vf^{\, \pe \pe}  \, ,
&&&&\cF_2^{\, \pe} \, = 5 \, \pr^{\, 4} \, \vf^{\, [3]} \, ,& 
\end{align}
where in particular the last one guarantees that, for spin $4$, $\cF_2^{\, \pe} = 0$.
This has the consequence that $\cA_{\vf}^{\, \pe}\, (a_1, a_2) = 0$, 
$\forall \, a_1, a_2$, and consequently \emph{any}
Lagrangian equation will be reduced on-shell to the form
\be \label{spin4}
\cF_2 \, - \, m^{\, 2} \, M_{\, \vf} \, = \, 0 \, ,
\ee
 It is not difficult to find that the right choice of $a$ and $b$ to 
guarantee that $\prd M_{\, \vf} = 0$ imply (\ref{fpagain}), together
with its consistency condition $\prd \vf^{\, \pe} \, = \, - \, \pr \, \vf^{\, \pe \pe}$
is
\be
M_{\, \vf} \, = \, \vf \, - \, \h \, \vf^{\, \pe} \, -  \, \h^{\, 2} \, \vf^{\, \pe \pe}\, .
\ee
Taking first a double trace and then a single trace of (\ref{spin4})
we find  in this way $\vf^{\, \pe \pe} =  0$ and then $\vf^{\, \pe} =  0$, which once again 
ensure that the Fierz system is recovered.

 \subsubsection{Spin $s$} \label{section2.2.3}

 In the general case, we look for a quadratic deformation
of the geometric Lagrangians  
giving rise to equations of motion 
of  the schematic form
\be \label{massiveeom}
\cE_{\, \vf} \, 
- \, m^{\, 2} \, M_{\vf} \, = \, 0 \, ,
\ee
where $\cE_{\, \vf}$ is a generic member in
the class of divergence-free Einstein tensors
recalled in Section \ref{section2.1.3}.
Again, all traces of $\vf$ are expected to contribute
to $M_{\vf}$, so that, for $s = 2\,n$ or $s = 2\, n + 1$, it
can generally be written as
\be \label{masstermb}
M_{\vf} \, =\, \vf \, + \, b_1 \, \h \, \vf^{\, \pe} \, + b_2 \, \h^{\, 2} \, \vf^{\, \pe \pe} \, 
+ \dots \, + \, b_k \, \h^{\, k}  \, \vf^{\, [k]} \, +  \dots  + \, b_n \, \h^{\, n} \,  \vf^{\, [n]} \, \, .
\ee
The same argument seen for spin $4$ applies also in this case:
we look for coefficients $b_1 \, \dots \, b_n$ such that $\cA_{\vf}^{\, \pe}$ 
vanishes on-shell, as a consequence of
$\prd M_{\, \vf} \, = \, 0$. Given  that 
no choice of the coefficients in $M_{\, \vf}$ exists such that 
$\prd M_{\, \vf} \, =\, 0$ implies $\cF^{\, \pe} \, = \, 0$ altogether, 
for the reasons discussed in the previous Section we are led 
to recover the Fierz-Pauli constraint (\ref{fpagain}), as a necessary condition
to relate the first two terms of $\cF^{\, \pe}$  with the
remainder of $\cA_{\, \vf}^{\, \pe}$.

 To this end we look for coefficents $b_1, \, \dots \, b_n$ such
that the divergence of (\ref{massiveeom}) imply (\ref{fpagain})
\emph{together with its consistency conditions}
\be
\prd \vf^{\, [k]} \, = \, - \, \fr{1}{2\, k \, - \, 1} \, \pr \, \vf^{\, [k + 1]} \, ,
\hspace{2cm} 
k \, = \, 1 \, \dots \, n \, .
\ee
By this we mean that, 
if we write the divergence of $M_{\vf}$ in the form
\be \label{divMb}
\prd M_{\vf} \, = \, \prd \vf \, + \, b_1 \, \pr \, \vf^{\, \pe} \, + \, 
\dots \, + \, \h^{\, k} \, (b_k \, \prd \vf^{\, [k]} \, + \, b_{k + 1} \, \pr \vf^{\, [k + 1]}) \, +  \, \dots \, ,
\ee
and we define
\be
\m_{\, \vf} \, \eq \, \prd \vf \, - \, \pr \, \vf^{\, \pe}\, , 
\ee
then we would like to rearrange (\ref{divMb}) as

\be \label{diverM}
\prd M_{\vf} \, = \, \m_{\vf} \, + \, \l_1 \, \h \, \m_{\vf}^{\, \pe} \, +
\, \dots \, + \, \l_k \, \h^{\, [k]}\, \m_{\vf}^{\, [k]} \, + \, \dots \, .
\ee

In this fashion, subsequent traces of (\ref{diverM}) would imply  
$\m_{\vf}^{\, [k]} = 0$, for $k = n, \, n-1 \, \dots$ and then finally $\m_{\vf} = 0$,
as desired\ft{This of course given that the coefficients $\l_k$ 
do not imply
any \emph{identical} cancellations among the 
traces of $\prd M_{\, \vf}$.}.
The form of $\m_{\vf}$ immediately fixes the first
coefficient to be $b_1  =  -  1$,
whereas consistency with (\ref{diverM})
requires
\be
\begin{split}
&\l_{\, k} \, = \, - \, \fr{b_{\, k}}{2\, k\, - \, 1} \, , \\
&b_{\, k\, + \, 1} \, = \, \fr{b_{\, k}}{2\, k\, - \, 1} \, ,
\end{split}
\ee
whose unique solution is 
\be
b_{\, k + 1} \, = \, - \, \fr{1}{(2 \, k \, - \, 1)\,! !} \, \,  . 
\ee
The following relations are then fulfilled, on-shell:
 
\begin{align} \label{onshellid}
& \cF \, = \, \Box \, \vf \, - \, \pr^{\, 2} \vf^{\, \pe} \, , &&&&\cF^{\, \pe} \, = 3 \, \pr^{\, 2} \, \vf^{\, \pe \pe}\, ,& \nonumber\\
&\cF_2 \, = \, \cF \, -\, 3\, \pr^{\, 4} \vf^{\, \pe \pe}  \, ,
&&&&\cF_2^{\, \pe} \, = 5 \, \pr^{\, 4} \, \vf^{\, [3]} \, ,& \nonumber\\
& \cF_3 \, = \, \cF_2 \, -\, 5 \, \pr^{\, 6} \, \vf^{\, [3]} \, ,
&&&&\cF_3^{\, \pe} \, = \, 7 \, \pr^{\, 6} \, \vf^{\, [4]} \, , \\
& . \, .\,  . &&&& .\, .\, . \nonumber\\
&\cF_n \, = \, \cF_{n - 1} \, -\, (2\, n \, - \, 1) \, \pr^{\, 2 n} \, \vf^{\, [n]} \,  ,
&&&&\cF_n^{\, \pe} \, = \, (2\, n \, + \, 1) \, \pr^{\, 2 n} \, \vf^{\, [n + 1]} \, . \nonumber 
\end{align}

The relevant point in this series of equations is that
the Fierz-Pauli constraint implies that $\cF_n^{\, \pe}  = 0$
for $s = 2 n , \, s = 2 n + 1$, and consequently
$\cA_{\vf}^{\, \pe} = 0$ for \emph{any}
Ricci tensor defined in (\ref{genlincomb}).
This means that \emph{any} Lagrangian equation  
reduces on-shell to the form
\be \label{reducedeom}
\cF_n \, - \, m^{\, 2} \, (\vf -  \h \, \vf^{\, \pe} \, - \, \cdots \, - 
\fr{1}{(2\, k \, - \, 3)\, !!} \, \h^{\, k} \, \vf^{\, [k]} \, \dots) \, = \, 0 \, .
\ee
As usual, subsequent traces of (\ref{reducedeom}) imply that
$\vf^{\, [k]} \, = \, 0 \, \, \forall k = n, n-1, \, \dots \, 1$. Finally
 the Fierz system is recovered by observing that, 
as a consequence of the recursion relation  
\be \label{recursF}
\cF_n \, = \, \Box \, \vf \, - \, \sum_{k = 1}^{n + 1} \, 
(2\, k \, - \, 1) \, \fr{\pr^{\, 2\, k}}{\Box^{\, k}}\, \vf^{\, [k]}\, , 
\ee
the tensors $\cF_n$ reduce to $\Box \vf$, once all traces 
of $\vf$ are set to zero.

 The final result is that consistent massive Lagrangians describing a spin-$s$ boson
are given by
\be \label{geomlags}
\cL \, = \,  \12 \, \vf \, \{\cE_{\, \vf} \, - \, 
m^{\, 2} \, M_{\, \vf} \} \, ,
\ee
with $\cE_{\vf}$ any of the Einstein tensors
constructed from the Ricci tensors (\ref{genlincomb}), and where
\be \label{mass}
M_{\, \vf} \, = \,  \vf -  \h \, \vf^{\, \pe}  -  \h^{\, 2} \, \vf^{ \, \pe \pe} 
- \fr{1}{3} \, \h^{\, 3} \, \vf^{ \, \pe \pe \pe} - \, \cdots \, - 
\fr{1}{(2\, k \, - \, 3)\, ! !} \, \h^{\, k} \, \vf^{\, [k]} \, - \,  \dots \, ,
\ee
is the generalised Fierz-Pauli mass term, for arbitrary integer spin.

%%%%%%%%%%%%%%%%%%%%%%%%%%%%%%%%%%%%%%%%%%%%%%%%%%%%%%%%%%%%%%%%%%

%%
\scs{Fermions} \label{section3}
 \subsection{Geometry for higher-spin fermions} \label{section3.1}
 \subsubsection{Fermionic curvatures} \label{section3.1.1}

 To describe fermions in a geometrical fashion, it is possible to reproduce 
the construction of the 
hierarchy of connections sketched in Section 
\ref{section2.1.1}, with the only modification
that the fundamental field 
\be
\ps_{\m_{s}}\,\eq \, \ps_{\m_1 \dots \m_s} \, ,
\ee
be understood as carrying 
a spinor index as well. We consider this field subject to the 
second of the transformations laws
(\ref{fierz}), that in symmetric notation reads
\be
\d \, \psi \, = \, \pr \, \e \, ,
\ee
but with an \emph{unconstrained} parameter $\e$.
 
 The whole construction then amounts to a  
rephrasing of the bosonic case \cite{dewfr}, the main result being 
that one can define for a rank-$s$ 
spinor-tensor\ft{In the following, whereas
this would not be source of confusion, and in order to
simplify the language, we shall refer to the
spinor-tensors $\psi$ loosely as ``tensors''.} the 
generalised connections
\be \label{gammaferm}
\G^{(m)}_{\, \m_{m}\, \n_{s}} =   \ \sum_{k=0}^{m}\fr{(-1)^{k}}{
\left(
{{m} \atop {k}}
\right)
}\ 
\pr^{\, m-k}_{\, \m}\pr^{\, k}_{\, \n} \, \ps_{\, \m_{k} \, \n_{s-k}}\, , 
\ee
whose gauge transformations are
\be \label{fermtrans}
\d\, \G^{(m)}_{\, \m_{m}\, \n_{s}} \, = \,(-1)^m \, (m+1)\, \pr^{\, m+1}_{\, \n} \, \e_{\, \m_{m},\, \n_{s-m-1}}\, . 
\ee
A fully gauge invariant tensor is first reached at the $s$-th step, and it is called 
a ``curvature" for fermionic gauge fields:
\be \label{fermcurv}
\cR_{\, \m_{s},\, \n_{s}} =   \ \sum_{k=0}^{s}\fr{(-1)^{k}}{
\left(
{{s} \atop {k}}
\right)
}\ 
\pr^{\, s-k}_{\, \m}\pr^{\, k}_{\, \n} \, \ps_{\, \m_{k} \, \n_{s-k}}\, .
\ee

\subsubsection{Generalised Dirac tensors} \label{section3.1.2}

 In analogy with the bosonic case, 
we would like to make use of the 
curvatures (\ref{fermcurv}) to construct generalised 
``Dirac-Rarita-Schwinger'' tensors sharing the symmetries of the
field $\psi$. If we insist that these tensors have the dimensions
of a first-order relativistic wave operator it is 
unavoidable to introduce non-localities, in the same fashion already
reviewed for bosons in Section \ref{section2.1.2}, where in particular  
``order-zero'' candidate Ricci tensors were
uniquely defined by eq. (\ref{kinetic}) requiring their degree of singularity
to be the lowest possible. In the fermionic case, however, more 
possibilities are allowed,  and a more refined analysis is needed to give an exhaustive
description of the linear theory, and in particular to
uncover the geometric meaning of the unconstrained equations
proposed in \cite{fs1, fs2}.

 From the technical viewpoint the basic novelty is that, while for bosons the only 
ways to saturate indices are
provided by traces and divergences, in the fermionic case we can also take
$\g$-traces.  In order to keep the degree of singularity
as low as possible, a first  definition of 
generalised Dirac tensors can then be obtained starting with the Ricci tensors 
(\ref{kinetic}), interpreted as functions of $\psi$, and replacing divergences with
$\g$-traces, while also formally acting with the operator
$\fr{{\not \pr}}{\Box}$ in the case of even rank, as summarised in the following table:
\begin{align} \label{diracgeom}
&Curvature& &spin& &\mbox{\emph{``Ricci''}} & && &spin& &\mbox{\emph{``Dirac''}} & \nonumber \\
&&&&&&&&&&&& \nonumber \\
& \cR_0 \, \sim \, \psi & & 0 & &\Box \, \cR_0 & &\rightarrow&
&1/2& &\fr{}{} \dsll \, \cR_0 & \equiv \, D_0 &\nonumber \\
& \cR_1 \, \sim \, \pr\, \psi & & 1 & &\fr{}{} \prd  \cR_1 & &\rightarrow&
&3/2& & \rsll_1 & \equiv \, D_1&\nonumber \\
& \cR_2 \, \sim \, \pr^{\, 2}\, \psi & & 2 & &\cR_2^{\, \pe} & &\rightarrow&
& 5/2 & & \fr{\dsll}{\Box}\, \cR_2^{\, \pe} &  \equiv \, D_2&\\
& \cR_3 \, \sim \, \pr^{\, 3}\, \psi & & 3 & &\fr{1}{\Box}\, \prd \cR_3^{\, \pe} & &\rightarrow&
& 7/2 & & \fr{1}{\Box}\, \rsll_3^{\, \pe} & \equiv \, D_3&\nonumber \\
& \cR_4 \, \sim \, \pr^{\, 4}\, \psi & & 4 & &\fr{1}{\Box}\, \cR_4^{\, \pe \pe} & &\rightarrow&
& 9/2 & & \fr{\dsll}{\Box^{\, 2}}\, \cR_4^{\, \pe \pe} & \equiv \, D_4& \nonumber \\
&& & \dots & && & \dots & && & \dots& \nonumber
%& \cR_s \, \sim \, \pr^{\, s}\, \psi & & s & &\fr{\prd^{s - 2[\fr{s}{2}]}}{\Box^{[\fr{s+1}{2}] - 1}}\, 
% \cR_s^{\, [\fr{s}{2}]} & &\rightarrow&
%& s + \12 & & (\fr{\dsll}{\Box^{\, 2}})^{s + 1 - 2 [\fr{s+1}{2}]}\, 
%\fr{(\g \cdot)^{s - 2[\fr{s}{2}]}}{\Box^{[\fr{s+1}{2}] - 1}}\, \cR_s^{\, [\fr{s}{2}]} \, \nonumber \\
\end{align}
The possibility of interchanging divergences and $\g-$traces
can also be used to replace one Lorentz-trace
with a $\g-$trace together with a divergence, according
to the formal substitution
\be
\h_{\, \m \n} \, = \, \12 \, \{\g_{\, \m}, \, \g_{\, \n}\} \, 
\ra \, \pr_{\, \m} \, \g_{\, \n} \, ,
\ee  
while still ensuring that
the total number of derivatives at the numerator be odd.
Whereas for the case of odd rank the 
corresponding tensors would be more singular than
the ones defined in (\ref{diracgeom}), and for this reason 
we neglect them as a first choice, in the case of
even rank only the minimum number of inverse powers of $\Box$ 
is needed to restore dimensions, and in this sense the tensors defined in this manner 
are \emph{a priori} equivalent candidates for the description of the 
dynamics:

\begin{align} \label{diracgeom2}
& \cR_2 \, \sim \, \pr^{\, 2}\, \psi & & 2 & &\cR_2^{\, \pe} & &\rightarrow&
& 5/2 & & \fr{1}{\Box}\, \prd \rsll_2 &  \equiv \, \hat{D}_2 &\nonumber \\
& \cR_4 \, \sim \, \pr^{\, 4}\, \psi & & 4 & &\fr{1}{\Box}\, \cR_4^{\, \pe \pe} & &\rightarrow&
& 9/2 & & \fr{1}{\Box^{\, 2}}\, \prd \rsll_4^{\, \pe} & \equiv \, \hat{D}_4&\\
&& & \dots & && & \dots & && & \dots& \nonumber \, 
%& \cR_s \, \sim \, \pr^{\, s}\, \psi & & s & &\fr{\prd^{s - 2[\fr{s}{2}]}}{\Box^{[\fr{s+1}{2}] - 1}}\, 
% \cR_s^{\, [\fr{s}{2}]} & &\rightarrow&
%& s + \12 & & (\fr{\dsll}{\Box^{\, 2}})^{s + 1 - 2 [\fr{s+1}{2}]}\, 
%\fr{(\g \cdot)^{s - 2[\fr{s}{2}]}}{\Box^{[\fr{s+1}{2}] - 1}}\, \cR_s^{\, [\fr{s}{2}]} \, \nonumber \\
\end{align}

 It is then clear that, in the fermionic case, 
keeping the singularity of the candidate tensors as
low as possible it is not a sharp enough criterion  to allow 
the identification of  
a \emph{unique} geometric theory.
Rather, when the rank of $\ps$ is even, say $2n$,
the most general candidate  has the 
form\ft{Barring a possible overall normalisation.}
\be \label{fermkintensab}
\cD_{2n} \, (a_{2n}) \, =  \, a_{2n} \, i \, D_{2n} \, + \, 
(1\,-\,a_{2n}) \, i \, \hat{D}_{2n} \, .
\ee

 We would like to clarify the meaning of this lack of uniqueness
in the definition of the basic tensors, and in particular 
to explain the role played in this context 
by the gauge-invariant, unconstrained, non local tensors proposed in \cite{fs1, fs2}, 
defined for spin $s = 2n +\fr{1}{2}$ and  $s=2n + \fr{3}{2}$ 
by the following recursion relations:
\be \label{fermkintens}
\cS_{\, n+1} \, = \, \cS_{\, n} +  \fr{1}{\, n\, (2n+1)}  \, \fr{\pr^{\, 2}}{\Box}\,
\cS_{\, n}^{\,  \pe}
\, - \, \fr{2}{2n+1} \, \fr{\pr}{\Box}  
\prd \cS_{\, n}\, ,  
\ee
where
\be
\cS_1 \, \eq \, \cS \, 
\ee
is the Fang-Fronsdal tensor (\ref{ffronsdalT}).

 Moreover, for the generalised tensors (\ref{fermkintensab})
one should also discuss the basic consistency issue of compatibility
with the Fang-Fronsdal theory, namely that the 
postulated equation of motion
\be \label{diraceqab}
\cD_{2n} (a_{2n}) \, = \, 0 \, ,
\ee
imply a compensator-like equation of the form
\be \label{compferm}
\cS \, = \, 2 \, i \, \pr^{\, 2}  \, \cK_{\psi}\, (a_{2n}) \, ,
\ee
where  
$\cK_{\psi}\, (a_{2n})$ should be a non-local tensor
shifthing as $\esl$ under the transformation $\d \, \psi \, = \, \pr \, \e$, 
in order to compensate the unconstrained
gauge variation of the Fang-Fronsdal tensor (\ref{ffronsdalT})
\be
\d \, \cS \, = \, - \, 2 \, i \, \pr^{\, 2} \, \esl \, .
\ee

 In the remainder of this Section we shall give an answer to the first
question, making  the geometrical meaning of (\ref{fermkintens}) explicit.
The main tool we shall resort to will be the comparison 
between the gauge transformations (\ref{fermtrans}) of the connections
defined in Section \ref{section3.1.1} and those of the kinetic 
tensors $\cS_{\, n + 1}$ defined in (\ref{fermkintens}) (which are not gauge-invariant, 
if $s \, > \,  2n + \fr{3}{2}$).

 In the next Section we shall discuss the role played by these tensors, 
under the criterion that (\ref{diraceqab})
be deducible from a Lagrangian. For the subclass
of tensors meeting this requirement it will be easy to show
consistency with the Fang-Fronsdal theory.

 To begin with, let us observe that no ambiguity manifests itself 
in the odd-rank case, where it is possible to show that  (\ref{diracgeom}) 
and  (\ref{fermkintens}) actually \emph{coincide}:
\be \label{diracodd}
\cS_{\, n} \, = \,i \,  D_{2n - 1}  \, .
\ee
In order to clarify this identity it is useful to compare 
the gauge transformations  (\ref{fermtrans})
of the de Wit-Freedman connections
with those of the tensors $\cS_n$,
\be
\d \, \cS_n \, = \, - \, 2 \, i \, n \, \fr{\pr^{\, 2n}}{\Box^{\, n-1}}\, \esl^{\, [n-1]}\, ,
\ee 
and to observe that, if the same gauge
transformation is implemented by one of the tensors $\cS_n$ 
and one connection $\G$, suitably modified in order for
indices and dimensions to match, we can 
infer that these two quantities actually define the same tensor. 
Indeed, they could only differ by gauge invariant quantities,
but \emph{by construction} the kinetic tensors, as well as the connections,
do not contain gauge-invariant ``sub-tensors''.

 This justifies the following identification:
\be \label{tabular}
\d\, \cS_n\, =\, \d \fr{i}{\Box^{n-1}}\, \gsl^{(2n-1)\, [n-1]} \ \ \  \ra \ \ \ 
\cS_n \,=\, \fr{i}{\Box^{n-1}}\, \gsl^{(2n-1)\, [n-1]}\, .
\ee
This last equality makes it clear that it is only in the odd-rank case
that the tensors (\ref{fermkintens}) can be given a 
straightforward geometric interpretation. Indeed,  considering 
the first value of $n$ such that the two tensors 
are gauge-invariant, (\ref{tabular}) automatically reduces to (\ref{diracodd}).

 On the other hand, for the even-rank case,
the kinetic tensors (\ref {fermkintens}) can be expressed as
linear combinations of the geometric ones defined in 
(\ref{diracgeom}) and (\ref{diracgeom2}).
For instance, in the case of spin $s = \fr{5}{2}$, 
writing all three tensors involved in terms of the 
Fang-Fronsdal tensor $\cS$, 
\be \label{52second}
\begin{split}
&i\, D_2 \, =\, \cS \, + \,  \fr{\pr^{\, 2}}{\Box}\, 
\cS^{\, \pe} \, - \, \fr{\pr}{\Box}\, \prd \cS \, , \\
& i\, \hat{D}_2 \, = \, \cS \, - \, \12 \, \fr{\pr}{\Box}\, \prd \cS \, , \\
&\cS_2 \, = \, \cS \, + \,  \fr{1}{3}\, \fr{\pr^{\, 2}}{\Box}\, \cS^{\, \pe} \, - 
\,\fr{2}{3}\,  \fr{\pr}{\Box}\, \prd \cS \, ,
\end{split}
\ee
it is  easy to realize that  $\cS_2$
is just a member of the class defined in (\ref{fermkintensab})
corresponding to the case $a_2 = 1/3$.
 
 To make the general validity of 
this observation explicit, let us recall  the recursive definition of the connections
\cite{dewfr}:

\be \label{dwfconn}
 \G^{(m)}_{\, \s \r_{m-1},\, \m_s} \, = \, \pr_{\, \s} \G^{(m-1)}_{\, \r_{m-1},\, \m_s}\, - \, 
                                   \fr{1}{m} \, \pr_{\, \m} \G^{(m-1)}_{\, \r_{m-1},\, \s\m_{s-1}} \, .
\ee

If $m\, = \,2n$, this last equality relates  $\G^{(2n)}$ to  $\G^{(2n-1)}$, which,
in its turn, can be related  to the tensors $\cS_n$, according to (\ref{tabular}).
As a consequence, the following equalities hold:

\begin{alignat}{4}
 &\fr{i}{\Box^{\, n}}\, \prd \gsl^{\, (2n)\, [n-1]}\ \  &= &\ \ \cS_n \, - \, \fr{1}{2n} \, \fr{\pr}{\Box} \, \prd \cS_n \, ,
 \label{geom1} \\
 &i\, \fr{\dsll}{\Box^{n}}\, \G^{(2n)\, [n]}\ \ &= &\ \  \cS_n \, - \, \fr{1}{2n} \, \fr{\pr}{\Box} \, \dsll \ssl_n \, .
 \label{geom2}
\end{alignat}
and using the generalised Bianchi identities verified
by $\cS_n$ \cite{fs1, fs2},
\be \label{nonlocbianchi}
\prd \cS_n \, - \, \fr{1}{2n} \, \pr \, \cS_n^{ \, \pe} \, 
 - \, \fr{1}{2n}\,  {\not {\! \pr}} \ssl_n 
\, = \, i \, \fr{\pr^{2n}}{\Box^{n-1}} \psisl^{[n]} \, ,
\ee
we can rewrite (\ref{geom2}) as follows:
\be \label{geomsec}
i \,\fr{\dsll}{\Box^{n}}\, \G^{(2n)\, [n]}\, = \, \cS_n \, - \,\fr{\pr}{\Box} \, \prd \cS_n \, + 
\, \fr{1}{n} \, \fr{\pr^{\, 2}}{\Box} \, \cS_n^{ \, \pe} \, + 
\, i \, (2n+1) \, \fr{\pr^{2n+1}}{\Box^{n}} \psisl^{[n]} \, .
\ee
Next,  looking for a combination of (\ref{geom1}) and (\ref{geomsec}) such
as to reproduce $\cS_{n+1}$, \emph{modulo} the term in $\psisl^{[n]}$,
we find

\be \label{rankeven}
\fr{i}{2\, n\, + \, 1} \, \left[ \fr{\dsll}{\Box^{n}}\, \G^{(2n)\, [n]}\right] \, 
+\, \fr{2\, n}{2\, n\, + \, 1} \, \left[\fr{i}{\Box^{n}}\, \prd \gsl^{(2n)\, [n-1]}\right]  \,= \, 
\cS_{n+1} \, + \, i \, \fr{\pr^{2n+1}}{\Box^{n}} \psisl^{[n]} \, .
\ee
If $s\, = \, 2\, n \, + \, \12$ then 
$ \psisl^{[n]}$ is not present, and the tensors on the 
l.h.s. just reproduce $D_{2n}$ and $\hat{D}_{2n}$
respectively; this proves that  
$\cS_{n \, + \, 1}$ is a linear combination of the 
form (\ref{fermkintensab}) according to
\be \label{rankeven}
\cS_{n \, + \, 1} \, = \, \fr{i}{2\, n\, + \, 1} \, D_{2n} \, + \, 
i \, \fr{2\, n}{2\, n\, + \, 1} \, \hat{D}_{2n} \, .
\ee

 To summarise, the definition of non-local, least singular, 
generalised Dirac tensors is unique only for odd-rank spinor-tensors, 
and is given by (\ref{diracodd}). For even-rank spinor-tensors we have 
in principle a one-parameter
family of candidates given by (\ref{fermkintensab}), and the
kinetic tensors (\ref{fermkintens}), first introduced in \cite{fs1, fs2},
are just one specific member in this family, as indicated by
(\ref{rankeven}).

 A higher degree of non-uniqueness could also be
considered if, in analogy with the discussion of the bosonic case
of Section \ref{section2.1.2}, we take into account the 
possibility of introducing more singular contributions, involving 
further $\g-$traces of the tensors given by  
(\ref{fermkintensab}) and (\ref{diracodd}). By including in the definition
of the $\cD_n \, (a_n)$ the odd-rank case, 
with the corresponding coefficient always to be chosen as
$a_{2n + 1} = 1$, we can define in general
the fermionic analogue of (\ref{genlincomb}) by
\be \label{fermgeneric}
\begin{split}
\cW_{\psi}\, (a_n \, ;\{c_k, d_k\}) \, = \, \cD_{n} \, (a_n) \, & + \, 
\fr{1}{\Box} \, (c_1 \, \pr \dsll \, {\not \! \! \cD}_{n} \, \, (a_n) \, + 
\, d_1 \, \pr^{\, 2} \, \cD_{n}^{\, \pe}\, (a_n)) \,+ \, \dots \\
& + \, \fr{1}{\Box^{\, k}}\, (c_k \, \pr^{\, 2k-1} \dsll \, {\not \! \! \cD}_{n}^{\, [k-1]} \, \, (a_n) \, 
+ \, d_k \, \pr^{\, 2k} \, \cD_{n}^{\, [k]}\, (a_n)) \,+ \, \dots \, .
\end{split}
\ee

 Finally, of course, the question remains whether one or more representatives
among the whole family of generalised Dirac tensors could play any special role. 
As we shall see in the next Section, 
the requirement that the equation
\be
\cD_{n} \, (a_{n}) \, = \, 0 \, ,
\ee
could be derived from a Lagrangian 
introduces a great simplification in the full description, 
but still does not imply the selection of 
a unique representative among the  $\cD_{n} \, (a_{n})$.

\subsubsection{Geometric Lagrangians} \label{section3.1.3}

 We now look for a Lagrangian derivation of the 
equation
\be \label{fermieq}
\cD_{n} \, (a_{n}) \, = \, 0 \, ,
\ee
with $\cD_{n} \, (a_{n}) $  defined in (\ref{fermkintensab}) for
even $n$, and in (\ref{diracodd}) for odd $n$, where
$n$ is the rank of $\psi$. 

 In the odd-rank case we already know the solution, since the Einstein
tensors for (\ref{diracodd}) were constructed in \cite{fs1, fs2}, and will be recalled later, 
In the even-rank case, the one in which
a true ambiguity exists, the final outcome
will be that only \emph{two} tensors, among the infinitely many
defined in (\ref{fermkintensab}),
can be used to write a gauge-invariant Lagrangian.

 For instance, in the first non-trivial case of spin $s = 5/2$,
if we try to construct a divergenceless Einstein
tensor from (\ref{fermkintensab}) in the form
\be
\cG_{a_2}\, (k,\, \l) \, = \, 
\cD_{2} \, (a_2) \, + \, k \, \g \, {\not \! \! \cD}_2 (a_2) \, +
\l \, \h \, \cD_2^{\, \pe} (a_2) \, ,
\ee
it is possible to verify that the condition
\be
\prd \cG_{a_2}\, (k,\, \l) \, \eq \, 0 \, ,
\ee
admits, together with the known solution 
given by \cite{fs1, fs2}
\be \label{spin52ferm}
\cG_{1/3}\, (-1/4,\, -1/4)\, = \, 
\cS_2 \, - \, \fr{1}{4} \, \g \, {\not \! \cS}_2  \, -
\fr{1}{4}  \, \h \, \cS_2^{\, \pe}  \, ,
\ee
 only
a second solution, namely
\be \label{spin52bose}
\cG_{1}\, (0,\, -1/2)\, = \, 
D_2 \, - \, \12 \, \h \, 
D_2^{\, \pe}\, .
\ee 
With hindsight, the existence of this second possibility is not
surprising, since the algebraic properties of
$ D_2 \, = \, \fr{\dsl}{\Box}\, \cR_2^{\, \pe}$ are the same 
as for the corresponding bosonic tensor, and then, in this case,
we could have expected to find the fermionic counterpart of
the linearised Einstein tensor of Gravity.

 Maybe it might be less clear what is the obstruction
for the other equations in (\ref{fermieq})
to be derived from a Lagrangian, but indeed there
is a simple algebraic reason, that can be
explained looking at the full general case.

In order to find solutions 
to the equation\ft{We reintroduce the label $2n$ to stress
that the following considerations refer to the even-rank case.}
\be \label{divferm}
\prd \cG_{a_{2n}}\, (\{k_i, \, \l_i\}) \, \eq \, 0 \, ,
\ee
where
\be
\cG_{a_{2n}}\,(\{k_i, \, \l_i\}) \, = \, 
\cD_{2n} \, (a_{2n}) \, + \, 
\sum_i \, \left[k_i \, \g  \, \h^{\, i-1} \, {\not \! \! \cD}_{2n}^{\, [i-1]} \, (a_{2n}) \, + \, 
\h^{\, i} \, \l_i \, \cD_{2n}^{\, [i]} \, (a_{2n}) \right] \, ,
\ee
we must look for coefficients  $a_{2n}, \, k_1, \, \dots, \, k_{\, n-1}, \l_1,\, \dots, \, \l_n$ such to
imply the needed chain of cancellations among equivalent
tensorial structures. On the other hand, in (\ref{divferm})
an isolated contribution will appear of the form
\be
\prd \cG_{a_{2n}}\,(\{k_i, \, \l_i\}) \, \sim \, \h^{\, n  - 1} \, 
\g \, \prd \, {\not \! \! \cD}_{2n}^{\, [n-1]}\,(a_{2n})  ,
\ee
that consequently must vanish identically, 
thus defining a linear equation in the coefficient 
$a_{2n}$ alone, that can admit at most one solution. This solution
can only correspond to the Einstein tensors 
generated from (\ref{fermkintens})
(since we know that such a  solution exists), that we report here
in the general case of any rank \cite{fs1, fs2}
\be \label{einstferm}
\cG_n \, = \, \cS_n \, + \,
\sum_{0 < p \leq n} \ \frac{(-1)^p}{2^p \ p! \ 
\left( {n \atop p} \right)} \, \eta^{p-1} \left[\  \eta \ 
{\cal S}_n^{\, [p]} \ + \ \gamma  \ {\cal {\not {\! S}}}_n^{\, [p-1]}
\ \right] \, .
\ee
The only possible exception to this argument
could be  that the term 
\be
\h^{\, n  - 1} \, 
\g \, {\not \! \! \cD}_{2n}^{\, [n-1]}\,(a_{2n})
\ee
is not  present at all in $\cG_{a_{2n}}\,(\{k_i, \, \l_i\})$, 
as it is the case if the coefficient $k_{n - 1}$ is
chosen to be zero. This is possible, but then the argument
can be iterated backwards, and leads to the conclusion
that \emph{all} $k_i$ coefficients  must vanish, that is to say
no bare $\g$'s should appear in $\cG_{a_{2n}}\,(\{k_i, \, \l_i\})$.
At that point  (\ref{divferm}) would be of the form
\be \label{diverG2}
\prd \cG_{a_{2n}}\, (\{k_i \, \eq \, 0, \, \l_i\}) \, = \, 
\prd \, \cD_{2n} \, + \, \l_1 \, \pr \, \cD_{2n}^{\, \pe}\, + \, 
\h\, (\l_1 \, \prd \, \cD_{2n}^{\, \pe} \, + \, \l_2 \, \prd \cD_{2n}^{\, \pe \pe}) \, + \, \dots \, ,
\ee
where it is important to notice that the two contributions
contained in $\cD_{2n}$ must undergo \emph{separate} cancellations.
Indeed, from (\ref{fermkintensab})
\be %\label{fermkintensab}
\cD_{2n} \, (a_{2n}) \, = \, a_{2n} \, D_{2n} \, + \, 
(1\,-\,a_{2n}) \, \hat{D}_{2n} \, ,
\ee
it is possible to appreciate that the bosonic-like contribution
given by $D_{2n}$ cannot be used to compensate 
terms in $\hat{D}_{2n} \sim \prd \, {\not \! \! \cR}_{\, 2 n}^{\,[n-1]}$, because
of the terms in ${\not \! \psi}$ only present in this 
second tensor.
This means that, for instance, the following 
cancellations should occur simultaneously, if the Einstein
tensor has to be divergenceless:
\be
\begin{split} \label{ruleout}
a_{2n} \{\prd \, D_{2n} \, + \, \l_1 \, \pr \, D_{2n}^{\, \pe}\} = \, 0 \, ,\\
(1 \, - \,  a_{2n} )\{\prd \, \hat{D}_{2n} \, + \, \l_1 \, \pr \, \hat{D}_{2n}^{\, \pe}\} = \, 0 \, ,
\end{split}
\ee
while more generally (\ref{diverG2}) splits into two series of independent
conditions, one for $D_{2n}$ and its traces, and
another one for $\hat{D}_{2n}$ and its traces.

 Now, whereas the first equation admits a known solution,
given by $\l_1 = - \fr{1}{2n}$ \cite{fs1, fs2},
(as can be deduced from the bosonic identities (\ref{bianchids})), 
it is possible to check that the second equation in (\ref{ruleout})
actually admits no solutions at all, as for instance can
be verified  explicitly if $s = \fr{5}{2}$, from the
expression of $\hat{D}_2$ given in (\ref{52second}).

 The conclusion of this analysis is that
the ambiguity in the definition of Dirac tensors
in the even-rank case contained in (\ref{fermkintensab})
actually persists at the level of the construction
of Lagrangians, but simplifies to only
two options. One option is given by the tensors
(\ref{fermkintens}), that also provide the unique solution
in the odd-rank case, and whose geometrical meaning
is encoded in (\ref{diracodd}) and (\ref{rankeven}).
The other possibility, which is competitive
with the first one only in the even-rank case, consists in simply
reinterpreting the bosonic field in (\ref{kinetbose}) as 
carrying a spinor index as well, thus getting (\ref{diracgeom2}).

 We expect these ambiguities, 
together with the full degeneracy
appearing when the more singular possibilities (\ref{fermgeneric}) are 
considered, to
disappear when the coupling
with an external current is turned on, in the same fashion discussed 
for bosons in \cite{FMS}.
Nonetheless, this analysis  has not
yet been performed, and is left
for future work.
 As we shall see, this will not prevent us
from analysing the mass deformation 
of the fermionic Lagrangian in its full generality.

 Before turning to the analysis of the massive case
we shall give an argument to show that the 
Lagrangian equations of this Section imply
the compensator equations (\ref{compferm}).
We do this for the equations given by $\cG_{\, n + 1} = 0$, 
with $\cG_{\, n}$ defined in (\ref{einstferm}).
The same argument, with minor modifications, also
applies to the other option described in this Section
(and to the bosonic tensors (\ref{kinetbose}) as well).
 
 The Lagrangian equations defined by (\ref{einstferm})
can be easily shown to imply $\cS_{\, n+1} = 0$.
This equation in turn, using (\ref{fermkintens})
and the Bianchi identity (\ref{nonlocbianchi}) can be 
cast it in the form\ft{The only relevant property of the coefficients, 
that we do not analyse in the following, is that they should not
imply \emph{identical} cancellations, at any stage in the
iterative procedure. This can be checked explicitly.}
\be \label{nonlocbianchi2}
\cS_n \, + \, a_n \, \fr{\pr}{\Box} \, \dsll \, {\not \! \! \cS}_n \, + \, b_n \, \fr{\pr^2}{\Box} \, 
\cS_n^{\, \pe}\, 
+ \, c_n \,  i \, \fr{\pr^{2n + 1}}{\Box^{n}} \psisl^{[n]} \, = \, 0 \, .
\ee
Computing the $\g$-trace of (\ref{nonlocbianchi2}) in order to express ${\not \! \! \cS}_n$ as the gradient
of a tensor, it is possible to rewrite it in the form 
\be
\cS_n \, =  \, 2 \, i \, \pr^2 \, \cK_n \, .
\ee
This procedure can be iterated by making repeated use of (\ref{fermkintens})
and (\ref{nonlocbianchi}).
For instance at the second step the result looks
\be
\cS_{n-1} \, = \, 2 \, i \, \pr^2 \, \{\cK_{n-1} \, + \, \cK_n \} \, ,
\ee
while after $n-1$ iterations one would find the desired expression
\be
\cS \, =  \, 2 \, i \, \pr^{\, 2} \, \cK_{\, \ps} \, ,
\ee
in which all non-localities are in the pure-gauge term $\cK_{\, \ps}$, and can then be eliminated
using the trace of the gauge parameter, reducing in this way the non-local dynamics
to the local Fang-Fronsdal form.

 Given the formal consistency of these theories at the free
massless level, in the  next Section we shall investigate the
possibility of finding a proper quadratic
deformation that would extend their meaning to the massive case.

\subsection{Mass deformation} \label{section3.2}

 In this Section we wish to reproduce and adapt to the fermionic case 
the  results discussed in Section \ref{section2.2} concerning
the massive phase of the bosonic theory. Since the basic ideas
and the methodology strictly resemble what we already
discussed in that context, here the presentation will be more concise.

 The Fierz-Pauli constraint (\ref{fpagain}) was the 
building-block of the full construction in the 
bosonic case. Hence,
the first piece of relevant information we have to obtain is about 
the analogous condition for fermions. 

 As was the case for the Proca theory, briefly
recalled in the Introduction, the 
analysis of its direct counterpart, the massive 
Rarita-Schwinger theory, only furnishes an incomplete 
information. For spin $\fr{3}{2}$ indeed, the massive deformation of the 
geometric Lagrangian is 
\be
\cL \, = \, \12 \, \bar{\psi} \, \{\cS \, - \, \12 \, \g \, \ssl \, + \, 
m \, (\psi \, - \, \g \, \psisl)\} \, + \, h.c. \, ,
\ee
whose equation of motion can be easily reduced 
to the Fierz system (\ref{fierzsyst}), \emph{via} 
the basic on-shell condition that it implies
\be \label{wrongfierz}
\prd \psi \, - \, \dsll \, \psisl \, = \, 0 \, .
\ee
Nonetheless, this constraint would not be
the correct one in the general case, and to uncover
the full Fierz-Pauli constraint for fermions 
it is necessary to analyse the example of spin $\fr{5}{2}$.

\subsubsection{Spin $5/2$} \label{section3.2.1}
 Let us consider the two different geometric formulation
defined by (\ref{spin52ferm}) and (\ref{spin52bose}). Starting
from (\ref{spin52ferm}), we write a tentative massive equation
in the generic form
\be \label{spin52mass}
\cS_2 \, - \, \fr{1}{4} \, \g \, {\not \! \cS}_2  \, -
\fr{1}{4}  \, \h \, \cS_2^{\, \pe}  \, - m \, 
(\psi \, + \, a \, \g \, \psisl \, + \, b \, \h \, \psi^{\, \pe}) \, = \, 0  ,
\ee
where we have included all possible terms in 
the definition of $M_{\, \psi}$. We look for 
coefficients $a$ and $b$ such that 
\be
\prd M_{\, \psi} = 0 \ \ \ra \ \ 
{\not \! \cS}_2 \, = \, 0 \, , 
\ee
which is a necessary conditions in order to recover the Fierz system 
(\ref{fierzsyst}).
From the explicit form  of $\ssl_{\, 2}$
\be
\ssl_{\, 2} \,  = \, \fr{4}{3} \, (\prd \psi \, - \, \dsll \, \psisl \, 
+ \, \fr{\pr}{\Box} \, \dsll \, \prd \psisl \, - \, \fr{\pr}{\Box} \, \prd \prd \psi) \, ,
\ee
we can see that actually if we could get (\ref{wrongfierz}) then 
$\ssl_{\, 2}$ would vanish. On the other hand, from the divergence
of the mass term
\be
\prd M_{\, \psi} \,(a, \,b) \, = \, \prd \psi \, + \, a \, \dsll \, \psisl \, 
+ \, a \, \g \, \prd \psisl \, + \, b \, \pr \, \psi^{\, \pe} \, ,
\ee 
we see that the only way to get rid of $\g \, \prd \psisl$ in this
expression is to show that it vanishes. The computation 
of the  $\g$-trace of 
$\prd M_{\, \psi} \,(a, \,b)$ 
\be
\g \, \cdot (\prd M_{\, \psi} \,(a, \,b) ) \, = \, [1 \, + \, a \, (D + 2)] \, \prd \psisl \, 
 + \, (b \, - \, a) \, \dsll \, \psi^{\, \pe}  \, = \, 0 \, ,
 \ee
makes it manifest that the desired condition is only achieved
if $a = b$, which already tells us that the mass term cannot have
the same form as in the spin-$\fr{3}{2}$ case. Some further manipulations
allow to conclude that to get rid of $\ssl_{\, 2}$ the following 
condition is needed:

\be \label{fpferm}
\prd \psi \, - \, \dsll \, \psisl \, - \, \pr \, \psi^{\, \pe} \, = \, 0 \, ,
\ee
which actually will represent the \emph{fermionic Fierz-Pauli constraint}
for any spin. The same condition
allows to write a consistent massive theory starting
from the alternative geometric option found in Section 
\ref{section3.1.2} and given for this case 
by the tensor $D_2$ in (\ref{diracgeom}).  Actually, it is possible to show that
(\ref{fpferm}) imply the following consequences
\be
\begin{split}
&i \, D_2 \, = \, i\, (\dsll \, \psi \, - \, \pr \, \psisl \, - \, 
\fr{\pr^{\, 2}}{\Box} \,\dsll \,\psi^{\, \pe}) \, , \\
&i \, {\not \! \! D}_2  \, =  \, i\, 3 \, (\fr{\pr^{\, 2}}{\Box} \, \dsll \psisl^{\, \pe} \, + \, 2 \, 
\fr{\pr^{\, 3}}{\Box} \, \psi^{\, \pe \pe}) \, .
\end{split}
\ee
It is simple to conclude that the consistency condition (\ref{fpferm}) 
can be obtained in this case by choosing the mass 
term in the form
\be
M_{\, \psi} \, = \, \psi \, - \, \g \, \psisl \, - \, \h \, \psi^{\, \pe} \, .
\ee

\subsubsection{Spin $7/2$} \label{section3.2.2}
 The basic observation, to be stressed once again, 
is that whereas generality suggests that all possible
$\g$-traces of the field enter the mass term $M_{\, \psi}$, 
this criterion should not be applied \emph{directly}
to the Fierz-Pauli constraint (\ref{fpferm}), that 
instead we want to be reproduced without modification.
The main reason which can be given at this level
is that otherwise it would not be possible to 
obtain the condition that  ${\not \! \cS}_{\, 2} = 0$
on-shell.
In order to find the proper generalisation of  $M_{\, \psi}$
consistent with  (\ref{fpferm}) we then to look for a combination
\be
M_{\, \psi} \, (a, \, b, \, c) \, = \, \psi \, + \, a \, \g \, \psisl \, + \, b \, \h \, \psi^{\, \pe}
\, +c \, \h \, \g \, \psisl^{\, \pe} \, ,
\ee
such that its divergence could be cast in the form
\be
\begin{split}
\prd M_{\, \psi} \, (a, \, b, \, c) \,& = \, 
\prd \psi \, - \, \dsll \, \psisl \, - \, \pr \, \psi^{\, \pe} \, \\
& +\, \l_1 \, \g \cdot (\prd \psi \, - \, \dsll \, \psisl \, - \, \pr \, \psi^{\, \pe} ) \, \\
& + \, \l_2 \, \h \, (\prd \psi \, - \, \dsll \, \psisl \, - \, \pr \, \psi^{\, \pe} )^{\, \pe} \, .
\end{split}
\ee
This will guarantee that the divergence of the Lagrangian equation
will produce (\ref{fpferm}) among its consequences. 
Since in the following Section we shall give a complete 
treatment of the general case, here we just 
report the result
\be
M_{\, \psi} \, = \, \psi \, - \, \g \, \psisl \, -  \, \h \, \psi^{\, \pe}
\,- \, \h \, \g \, \psisl^{\, \pe} \, .
\ee

\subsubsection{Spin $s + 1/2$} \label{section3.2.3}
 In the general case, in order to derive the Fierz system 
(\ref{fierzsyst}) from 
a quadratic deformation of the geometric
Lagrangians, we look for a linear combination 
$M_{\, \psi}$ of $\g$-traces of $\psi$ such that,
starting from 
\be
\cL \, = \, \psb \, \{ \cE_{\, \psi}\, - \, m \, M_{\, \psi}\} \, + \, h.c. \, ,
\ee
the divergence of the corresponding equations of motion will imply
the \emph{Fierz-Pauli constraint} for fermions
\be \label{fpfermi}
\m_{\, \psi} \, \eq \, \prd \psi \, - \, \dsll \, \psisl \, - \, \pr \, \psi^{\, \pe} \, = \, 0 \, ,
\ee 
together with its consistency conditions
\begin{alignat}{4}
&{\not \! \m}_{\, \psi}^{\, [n]} \, & = & \, - \, [(2\, n \, + \, 1) \, \prd \psisl^{\, [n]}
 \, + \, \pr \, \psisl^{\, [n + 1]}] \, = \, 0 \, , \\
&\mp^{\, [n]} \, & = & \, - \, [(2\, n \, - \, 1) \, \prd \psi^{\, [n]}
\, + \, \dsll \, \psisl^{\, [n]} \, + \, \pr \, \psi^{\, [n + 1]}] \, = \, 0 \, . 
\end{alignat}
Let us consider then the general linear combination of $\g$-traces of $\psi$ 
written in the form
\be
M_{\, \psi} \, = \, \psi \, - \, \sum_{j = 0}^{[\fr{s -1}{2}]} \, a_{2j +1}
\, \g \, \h^{\, j} \, \psisl^{\, [j]} \, - \, 
\sum_{i = 1}^{[\fr{s}{2}]} \, a_{2 i}
\, \h^{\, i} \, \psi^{\, [i]} \, ,
\ee
and let us compute its divergence
\be
\begin{split}
\prd M_{\, \psi} \, = \, \prd \psi \,& - \, \sum_{j = 0}^{[\fr{s -1}{2}]} \, a_{2j +1}
\, \{\dsll \, \h^{\, j} \, \psisl^{\, [j]} \, + \,
\g \, \h^{\, j - 1} \, \pr \, \psisl^{\, [j]} \, + \,
\, \g \, \h^{\, j} \, \prd \, \psisl^{\, [j]}\} \\
 & - \, \sum_{i = 1}^{[\fr{s}{2}]} \, a_{2 i} \, \{
\, \h^{\, i - 1} \, \pr \, \psi^{\, [i]} \, + \, \h^{\, i} \, \prd \psi^{\, [i]}\} \, .
\end{split}
\ee
In order for it to be rearranged as
\be
\prd M_{\, \psi} \, = \, \mp \, + \, \l_1 \, \g \, \mpsl \, + \, 
\l_2 \, \h \, \mp^{\, \pe} \, + \, \dots \, 
+\l_{2 k} \, \h^{\, k} \, \mp^{\, [k]} \, 
+ \,  \l_{2 k + 1} \, \g \, \h^{\, k} \, \mpsl^{\, [k]} \, \dots \, .
\ee
the solution for the coefficient $a_k$ is unique and has the form
\be
\begin{split}
& a_{2 k + 2} \, = \, \fr{1}{(2 k - 1)! !} \, \\
& a_{2 k + 3} \, = \, \fr{1}{(2 k + 1)! !} \, ,
\end{split}
\ee
so that the generalised Fierz-Pauli mass-term for fermions 
is\ft{Just for the sake of keeping the formula compact, we are 
defining here $(- 1)!! = 1$.}

\be \label{massferm}
M_{\, \psi} \, = \, \psi \, - \, \sum_{j = 0}^{[\fr{s -1}{2}]} \, \fr{1}{(2 j - 1)! !}
\, \g \, \h^{\, j} \, \psisl^{\, [j]} \, - \, 
\sum_{i = 1}^{[\fr{s}{2}]} \, \fr{1}{(2 i - 3)! !}
\, \h^{\, i} \, \psi^{\, [i]} \, .
\ee

Again, the equation $\prd M_{\, \psi} \, = \, 0$
implies that all $\g$-traces of $\mp$, and then 
$\mp$ itself, vanish on-shell. This in turn
implies the following consequences

\begin{align} \label{onshferm}
& \cS_1 \, = \, i\, (\dsll \, \psi \, - \, \pr \, \psisl) \, , &
&\ssl_1  \, =  \, - \, i  \, \pr \, \psi^{\, \pe} \, , \nonumber\\
& \cS_2 \, = \, \cS_1 \, - \, i (\fr{\pr^{\, 2}}{\Box} \,\dsll \,\psi^{\, \pe} \, +\,
3\, \fr{\pr^{\, 3}}{\Box} \,\psisl^{\, \pe}) \, ,
&&\ssl_2  \,=\, 3\, i\, \fr{\pr^{\, 3}}{\Box} \, \psisl^{\, \pe \pe} \, , \nonumber\\
& . \, .\,  . && .\, .\, . \, , \\
&\cS_{n + 1} = \cS_n - i \{ (2 n - 1)
\fr{\pr^{\, 2 \, n}}{\Box^{\, n}} \dsll \,\psi^{\, [n]} +
(2 n + 1)\, \fr{\pr^{\, 2 n \, + 1}}{\Box^{\, n}} \psisl^{\, [n]}\} \, ,
&&\ssl_{n + 1} =  i (2 n + 1)  \fr{\pr^{\, 2  n + 1}}{\Box^{\, n}} 
\psi^{\, [n + 1]} \, . \nonumber
\end{align}

In particular from the equation for $\ssl_{n + 1}$ we deduce that
\emph{any} Ricci tensor of the form
\be
\cS_{n + 1}\, + \, b_1 \, \g \, \ssl_{n + 1} \, + \, \dots \, , 
\ee
reduces to $\cS_{n + 1}$, for $s = 2n + \12$ and $s = 2n + \fr{3}{2}$, and then
the Lagrangian equations of motion 
\be
\cE_{\, \psi} \, - \, m \, M_{\, \psi} \, = \, 0 \, ,
\ee
reduce to
\be
\cS_{n + 1} \, - \, m \, M_{\, \psi} \, = \, 0 \, .
\ee
As usual, successive traces of this equation allow to conclude 
that all $\g$-traces of $\psi$ vanish on-shell, which 
in turn implies, because of the recursive relations 
(\ref{onshferm}), that $\cS_{n + 1} \, = \, \dsll \, \psi$, and the Fierz system
(\ref{fierzsyst}) is finally recovered.

 As already stressed for 
the example of spin $s= 5/2$, no conceptual differences are present in the construction
of the massive theory for the tensors $D_{\, 2n}$ defined in 
(\ref{diracgeom}), which were shown in Section \ref{section3.1.3}
to constitute an alternative possibility for the formulation
of the dynamics in the even-rank case.  So for instance, in the
spin $s = 9/2$ case
the Fierz-Pauli constraint implies
the following analogues of the relations given in table (\ref{onshferm}):
\begin{align} \label{onshferm2}
& D_4 \, = \, D_2 \, + \, i\, (- \, 3 \, \fr{\pr^{\, 3}}{\Box} \psisl^{\, \pe}
\, + \, 10 \, \fr{\pr^{\, 5}}{\Box^{\, 2}} \psisl^{\, \pe \pe} \, 
+ \, 3 \, \fr{\pr^{\, 4}}{\Box^{\, 2}} \, \dsll \, \psi^{\, \pe \pe})\, , &
&{\not \! \! D}_4  \, =  \, 5 \, i  \, \fr{\pr^{\, 4}}{\Box^{\, 2}} \, \dsll \, \psisl^{\, \pe \pe} \, ,
\end{align}
showing that, under the constraint (\ref{fpfermi}), the trace
of $D_4$ vanishes, and the usual argument leading to 
(\ref{fierzsyst}) can be applied.

 If, on the one hand, this observation might be taken as an
argument in favour of the correctness of this kind of 
description of the massive theory, it must be also admitted 
that it raises for fermions 
an issue of uniqueness, of the same kind as the one already observed
for bosons. 

 We shall propose an interpretation of this open point in the following Section, 
focusing on the bosonic case.

\scs{The issue of uniqueness} \label{section4}
\subsection{Setting of the problem}

 The construction of geometric theories for higher-spin fields
proposed in \cite{fs1, fs2, FMS}, and reviewed in Sections 
$2.1.2$ and $2.1.3$ does not produce a unique answer, 
in the sense that infinitely many gauge-invariant Lagrangians are
actually available, whose corresponding free equations can all be shown 
to imply the (Fang-)Fronsdal ones, after a suitable, partial gauge-fixing
is performed.  

 Nonetheless, in \cite{FMS} it was shown that 
the analysis of the on-shell 
behavior of the geometric theories was biased by the absence of couplings.
Indeed, turning on even a non-dynamical source, and thus
performing a deeper check of the consistency of those Lagrangians
against the structure of the 
corresponding propagators, allows 
to restore uniqueness, as only one theory
was proved to survive this most stringent test.

 For the massive Lagrangians proposed in this work
a similar issue is to be discussed. Indeed, even if the 
generalised mass terms  found by imposing consistency
with the Fierz-Pauli constraint are unique, still they can 
be used to describe the mass deformations
of \emph{any} of the geometric theories available at the free level,
providing massive Lagrangians all implying the Fierz systems
(\ref{fierzsyst}), on-shell.

 We thus wonder whether other criteria might
suggest a selection principle among those theories, 
and in particular what is the memory, if any, that
the massive deformation keeps of the unique
Einstein tensor  selected at the massless level.

 The most obvious thing to try would be to see whether,
even in this case, turning on couplings with external sources
could give any indications about the existence of some
``preferred'' choice. To show that this is \emph{not} the case, 
it is sufficient to analyse the example of spin $4$. 
Let us consider the massive, geometric theory for this case, in the
presence of an external, conserved current $\cJ$:
\be 
\cL \, = \,  \12 \, \vf \, \{\cE_{\, \vf} \, - \, 
m^{\, 2} \, (\vf -  \h \, \vf^{\, \pe}  -  \h^{\, 2} \, \vf^{ \, \pe \pe})\, \}\, - \, \vf \, \cdot \, \cJ \, ,
\ee
where $\cE_{\, \vf}$ is a generic member in the class of 
geometric Einstein tensors recalled in Section $2.1.3$.
The key point is that conservation of currents guarantees
that the equations of motion still imply
the Fierz-Pauli constraint (\ref{fpagain}), and then, 
as discussed in Section $2.2.2$ for spin $4$, 
\emph{any} Lagrangian equation
will reduce on-shell to the form 
\be \label{currenteq}
\cF_{\, 2} \, - \, m^{\, 2} \, (\vf -  \h \, \vf^{\, \pe}  -  \h^{\, 2} \, \vf^{ \, \pe \pe}) \, = \, \cJ \, ,
\ee
with $\cF_{\, 2} $ defined in (\ref{kinetic3,4}).
Moreover, again because of (\ref{fpagain}), the structure of 
$\cF_{\, 2}$ considerably simplifies to 
(\ref{recursF})
\be
\cF_{\, 2} \, = \, \Box \, \vf \, - \, \pr^{\, 2} \, \vf^{\, \pe} \, 
- \, 3 \, \pr^{\, 4} \, \vf^{\, \pe \pe} \, ,
\ee
where it is to be noted that, in the computation
of the current exchange, both the contributions
in $\pr^{\, 2} \, \vf^{\, \pe}$ and $\pr^{\, 4} \, \vf^{\, \pe \pe}$
vanish when contracted with a conserved $\cJ$. 

 This observation, which is easily generalised to all spins
by means of (\ref{onshellid}) (and (\ref{onshferm}), if we wish 
to apply the same argument to 
fermions), 
implies that, from the viewpoint of the coupling with 
conserved sources, the full structure of the 
geometric part of the equations of motion 
kind of \emph{disappears} behind the mass term $M_{\, \vf}$, the only 
remaining contribution of this sector of the Lagrangian being the term in $\Box \vf$. 

 On the other hand this means that, in the computation
of the current exchange, \emph{the full responsibility 
of giving the correct propagator is now in the structure of 
the mass term} $M_{\, \vf}$, thus providing a non-trivial
consistency check of its validity, coefficient by coefficient, so to speak.
This test gives a positive answer, since for example for the case 
of spin $4$ the result is\ft{The corresponding  computation 
for the massless case, using the correct theory, gives
\be
\cJ \cdot  \vf \, =  \fr{1}{p^{\, 2}} \, \{\cJ \cdot \cJ \, - \, 
\fr{6}{D + 2} \, \cJ^{\, \pe} \cdot \cJ^{\, \pe} \, + \,
\fr{3}{D (D + 2)} \, \ \cJ^{\, \pe \pe}\cdot \cJ^{\, \pe \pe} \}
\ee
thus showing the generalisation of the 
Van Dam-Veltman-Zakharov discontinuity \cite{vdvz},
already noticed in \cite{FMS}.
}
\be
\cJ \cdot  \vf \, =  \fr{1}{p^{\, 2} \, - \, m^{\, 2}} \, \{\cJ \cdot \cJ \, - \, 
\fr{6}{D + 3} \, \cJ^{\, \pe} \cdot \cJ^{\, \pe} \, + \,
\fr{3}{(D + 1) (D + 3)} \, \cJ^{\, \pe \pe}\cdot \cJ^{\, \pe \pe} \}\, ,
\ee
whose correctness can be checked by comparison with the corresponding
computation performed in the local setting \cite{FMS}. 
However, for the same reasons,
the coupling with external sources in the massive case does not yield
any indications at all on the existence of a possible preferred 
theory.

 A different criterion might be suggested by the analogy 
with the example of spin $2$. 

 We have already observed, at the beginning of Section $2.2$, that
the origin of the Fierz-Pauli mass term  can be traced back
to the Kaluza-Klein reduction of the massless theory from
$D+1$ to $D$ dimensions, the very form of the mass term itself being 
simply encoded in the coefficient of the D'Alembertian operator
in the Einstein tensor,
\be \label{quasifp}
\cR \, - \, \12 \, \h \, \cR^{\, \pe} \, \sim \, \Box \, (h \, - \, \h \, h^{\, \pe}) 
\, + \, \dots \,  .
\ee
On the other hand, it is worth stressing again that, 
once the mass term is fixed to have the Fierz-Pauli form, the structure of the 
``ancestor'' geometric theory stays ``hidden'' behind it,
in the sense that neither
the free equations of motion, nor the computation of
the current exchange actually allow to keep memory of it. 
Thus, for instance, we could consider the following 
non-local equation for the description of the massive
graviton
\be
\cR \, - \, \12 \, \h \, \cR^{\, \pe}  \, + \, a \, (\h \, - \, 
\fr{\pr^{\, 2}}{\Box}) \, \cR^{\, \pe} \, - \, m^{\, 2} \, (h \, - \, \h \, h^{\, \pe}) \,= \, 0 \, ,  
\ee
whose geometric part is still in terms of a divergenceless 
(and gauge-invariant, at the massless level) tensor, which
would clearly  describe an inconsistent massless theory, 
to begin with since the current exchange would give
in this case a  wrong result, for generic real $a$.

 Notwithstanding this deficiency, whenever this ``massive graviton'' 
is coupled to a conserved current, 
the implementation of the Fierz-Pauli constraint would still imply
that the Ricci scalar vanishes on-shell, and the computation of the 
massive propagator would then furnish the correct result. 

 To summarise, for spin $2$ (but not only for 
spin $2$) in the massive case the basic information is encoded
in the mass term, whose role mainly is to guarantee that
the Fierz-Pauli constraint (\ref{fpagain}) be enforced on-shell. This constraint,
on the other hand, is strong enough to obscure the detailed
structure of the massless sector of the Lagrangian, that appears
from this viewpoint of relative importance. Nonetheless, 
a clear link between the massive theory and the correct massless one
can be traced back to the structure of the coefficient of the 
D'Alembertian operator in the Einstein tensor of the proper
geometric theory, which is ultimately responsible for
the form of the mass term, upon Kaluza-Klein
reduction from $D+1$ to $D$ dimensions.

 Even if it is not straightforward to establish such a direct link
in the non-local setting, because of the inverse powers of
$\fr{1}{\Box} \ \ra \ \fr{1}{\Box - m^{\, 2}}$ that would 
appear in the reduction of the non-local theory, 
but that are not present in our construction, 
we are anyhow led to conjecture that 
the generalised Fierz-Pauli mass term (\ref{mass}) bears a 
direct relationship with the correct geometric 
theory synthetically described by eqs.  (\ref{einstein}) and (\ref{Aident}), of which
it should simply represent the coefficient of the 
D'Alembertian operator.

 We would like to stress that, since the mass
terms (\ref{mass}) and (\ref{massferm}) proposed in this work have been found
following a path completely independent of
any detailed knowledge of the underlying geometric theory, 
we think it is fair to say that 
to verify this conjecture would represent a robust check
of the internal consistency of the whole construction.

 In the following Section we shall build the setup to
quantitatively discuss this conjecture. This will allow us to take a closer look
at the structure of the explicit solution to the identities
(\ref{Aident}), and to check  our hypothesis
for the first few cases.
Anyway, even if we regard the support provided from these explicit computations
as a strong indication of its validity, 
still we have not yet a proof of the conjecture in its full generality.

\subsection{Testing the uniqueness conjecture}

 We would like to show that the coefficient of the naked 
D'Alembertian in the Einstein tensor (\ref{einstein}) 
has the same form of the
generalised mass term (\ref{mass}). Namely
\be \label{conj1}
\cA_{\, \vf} \, - \, \12 \, \h \, \cA_{\, \vf}^{\, \pe} \, + \, \h^{\, 2} \, 
\cB_{\, \vf} \, \sim \, \Box \, M_{\, \vf} + \, \dots \, , 
\ee
where 
\be
M_{\, \vf} \, = \, \vf \, - \, \h \, \vf^{\, \pe} \, - \, \h^{\, 2}
\, \vf^{\, \pe \pe} \, \dots \, - \, \fr{1}{(2k - 3) ! !} \, \h^{\, k} \, \vf^{\, [k]}\, + \, 
\dots \, .
\ee
We recall that $\cA_{\, \vf}$ has the compensator structure (\ref{G}), 
\be
\cA_{\, \vf} \, = \, \cF \, - \, 3 \, \pr^{\, 3} \, \g_{\, \vf} \, ,
\ee
and satisfies the two identities (\ref{Aident}), that we report here 
for simplicity,
\be \label{Aiden}
\begin{split}
& \prd \cA_{\, \vf} \, - \, \12 \, \pr \, \cA_{\, \vf}^{\, \pe}  \, \eq \, 0 \, , \\
& \cA_{\, \vf}^{\, \pe \pe}  \, \eq \, 0 \, .
\end{split}
\ee
More details on the explicit solution for $\cA_{\, \vf}$ are given in Appendix \ref{B}.

  The tensor $\cB_{\, \vf}$, defined as the solution to
the equation 
\be \label{meaningB}
\prd \{\cA_{\, \vf} \, - \, \12 \, \h \, \cA_{\, \vf}^{\, \pe} \, + \, \h^{\, 2} \, 
\cB_{\, \vf}\} \, \eq \, 0 \, .
\ee
can be decomposed in terms of a sequence of the form \cite{FMS}
\be
\cB \, = \, \cB_0 \, + \, \h \, \cB_1 \,+ \, \dots \, + \, 
\h^{\, k} \, \cB_k \, + \, \dots \, \h^{\, p} \, \cB_p \, ,
\ee
with 
$$
p \, = \, \mbox{integer part of } \{ \fr{s\, - \, 4}{2} \}\, \eq \, [ \fr{p - 4}{2} ] \, .
$$
In this way the condition implied by (\ref{meaningB}) can thus be turned 
into the system
\be \label{tensorsB}
\begin{split}
 \pr \, \cB_0 \, & = \, \12 \, \prd \cA_{\, \vf}^{\, \pe} \, , \\
 \dots &\,  , \\ 
 \pr \, \cB_k \, & = \, - \, \fr{k}{k \, + \, 2} \, \prd \, \cB_{\, k-1} \, , \\
 \dots & \, . 
\end{split}
\ee 
The double tracelessness of 
$\cA_{\, \vf}$ allows in this way to deduce from the first of (\ref{tensorsB}) the set of identities
\be
\prd \cB_0^{\, [k]} \, = \, - \, \fr{1}{2\, (k \, + \, 1)} \, \pr \, \cB_0^{\, [k+1]} \, ,
\ee
which, in turn, give the following relations among the tensors $\cB_k$ and the traces of 
$\cB_0$:
\be \label{tracesbk}
\cB_k \, = \, \fr{1}{2^{\, k - 1} \, (k \, + \, 2)!} \, \cB_0^{\, [k]} \, \, .
\ee
These relations, together with the explicit 
solution\ft{We are correcting here a misprint in the 
corresponding formula, ($4.69$), in \cite{FMS}.} for $\cB_0$, 
\be \label{explbzero}
\cB_0 \, = \, \12 \, \sum_{k=0}^{\, n-1} \,\fr{1}{2 k +  1}
\, \{a_k \, \fr{2\, k + 1}{2\, (n  -  k)} \, \fr{n  +  1 + k}{n + 1 - k} \, + \, 
a_{k + 1} \, \fr{n + 4 k  +  5}{2 (n  -  k)} \, + \, a_{k + 2} \}
\pr^{\, 2 k } \, \cF^{\, [k + 2]}_{\, n + 1} \, ,
\ee
allow to complete the construction of the Einstein tensor 
in the non-local, geometric case \cite{FMS}.

 Let us finally notice that, using (\ref{tracesbk}), it is 
possible to write the conjectured equality (\ref{conj1})
in the more explicit form
\be \label{conj2}
\cA_{\, \vf} \, - \, \12 \, \h \, \cA_{\, \vf}^{\, \pe} \,  +
\, \sum_{k = 0}^{[\fr{s - 4}{2}]} \fr{1}{2^{\, k}\, k!} \, \h^{\, k + 2} \, \cB_0^{\, [k]} \, 
\sim \, \Box \{\vf \, - \, \h \, \vf^{\, \pe} \, - \,
\sum_{k=0}^{[\fr{s - 4}{2}]} \, \fr{1}{(2\, k \, + \,1) ! !} \, \h^{\, k + 2} \, \vf^{\, [k + 2]}\} \, ,
\ee
where $[\fr{s - 4}{2}] = n - 1$ or $n - 2$ depending on whether
the spin is $s = 2n + 2$ or $s = 2n + 1$ respectively.

It is then clear from the form of $\cA_{\, \vf}$ that the
first two terms of $M_{\, \vf}$ are correctly reproduced by
$\cA_{\, \vf} \, - \, \12 \, \h \, \cA_{\, \vf}^{\, \pe}$, 
and the non-trivial part of the calculation is to check
whether the traces of $\cB_0$ satisfy
\be \label{tobeproven}
\cB_0^{\, [k]} \, = \, - \, \fr{2^{\, k}\, k!}{(2\, k \, + \,1) ! !} \, \Box \, \vf^{\, [k + 2]} \, + \, \dots \, .
\ee

 Since we have no explicit formula for the tensors
$\cF_{\, n + 1}$ in terms of $\cF$ or $\vf$, 
in order to compute the contribution of $\Box \, \vf^{\, [k + 2]}$ in
$\cB_0^{\, [k]}$ we shall not make use of (\ref{explbzero}). Rather, 
we shall exploit the results collected in Appendix  \ref{B}, 
about the structure of the tensor $\g_{\, \vf}$ in 
$\cA_{\, \vf}$ which are relevant to the present calculation.

 We would like to stress that we shall not 
attempt here to uncover the \emph{full} structure of $\cB_0$.  
Rather, keeping in mind our present goal, we shall
systematically discard contributions involving
\emph{divergences} of the field $\vf$ and of its traces, that
in our formulas will be collectively gathered under the label
``$irr$'', to indicate that they are irrelevant for the present 
purpose.

\subsubsection{Evaluation of $\cB_0^{\, [k]}$}
 The defining equation for $\cB_0$ is the first of 
(\ref{tensorsB}), that can be written more explicitly as
\be \label{bzero1}
2\, \pr \, \cB_0 \, = \, \prd \cF^{\, \pe} \, - \, 3 \, (\g_{\, \vf}
\, + \, 3 \, \pr \, \prd \, \g_{\, \vf} \, + \, 2 \, \pr^{\, 2} \, \prd \prd \g_{\, \vf} \, 
+ \, \pr^{\, 2}\, \g^{\, \pe}_{\, \vf} \, + \, \pr^{\, 3} \, \prd \, \g^{\, \pe}_{\, \vf}) \, ,
\ee 
where the general form of $\g_{\, \vf}$, given in eq. (\ref{structures})
in Appendix \ref{B}, together with the explicit knowledge
of its first coefficient, as can be read from (\ref{explicitgammas}), 
allow to express the first two terms in (\ref{bzero1})
as the gradient of a tensor
\be \label{graddelta}
\prd \cF^{\, \pe} \, - \, 3 \, \g_{\, \vf} \, = \, \pr \, \D \, , 
\ee
where
\be \label{delta}
\D \, = \, - \, 3 \, \sum_{q, \, l, \, m} \, \fr{a_{q l m}}{l} \, 
\pr^{\, l - 1} \, \prd^{ \, m} \, \cF^{\, [q]} \, .
\ee
with range of variation of the indices in the sum given in 
(\ref{rangeklm}).
A further simplification comes from the observation that
the basic identity (\ref{bident2}) satisfied by $\g_{\, \vf}$
can be also read as
\be
\prd \, \g_{\, \vf} \, = \, \fr{1}{4} \, (\vf^{\, \pe \pe} \, 
- \, \pr \, \g^{\, \pe}) \, ,
\ee
which, in turn, implies the set of relations
\be \label{tracesgamma}
\prd \, \g_{\, \vf}^{\, [k]} \, = \, \fr{1}{2\,(k + 2)} \, (\vf^{\, [k + 2]} \, 
- \, \pr \, \g^{\, [k + 1]}) \, .
\ee
As a consequence of (\ref{delta}) and (\ref{tracesgamma})
the equations for $\cB_0$ can be written in 
the following, relatively simple, form
\be
 2 \, \cB_0 \, = \, \D \, - \, \fr{9}{4} \, \vf^{\, \pe \pe} \, + \, 
\fr{1}{12} \, \pr^{\, 2} \, \vf^{\, [3]} \, + \, \fr{3}{2} \, 
\pr \, \g_{\, \vf}^{\, \pe} \, - \, \fr{1}{4} \, \pr^{\, 3} \, \g_{\, \vf}^{\, \pe \pe} \, + \, irr \, , 
\ee
whereas for the $k$-th trace of $\cB_0$ one finds

\be \label{bzero2}
2 \, \cB_0^{\, [k]} \, = \, \D^{\, [k]} \, + \, a_k \, \vf^{\, [k + 2]} \, + \, 
b_k \, \pr^{\, 2} \, \vf^{\, [k + 3]} \, + \, c_k \, 
\pr \, \g_{\, \vf}^{\, [k + 1]} \, + \, d_k \, \pr^{\, 3} \, \g_{\, \vf}^{\, [k + 2]} \, + \, irr \, ,
\ee

with coefficients $a_k$ - $d_k$ recursively defined by the following system

\be \label{systemakdk}
\begin{split}
a_{k + 1} \, & =  a_k \, + \, b_k \, + \, c_k \, \fr{1}{k + 3}\, , \\
b_{k + 1} \, & =  b_k \, + \, \fr{1}{k + 4} \, d_k \, , \\
c_{k + 1} \, & =  c_k \, \fr{k + 2}{k + 3} \, + \, d_k \, , \\
d_{k + 1} \, & =  d_k \, \fr{k + 1}{k + 4} \, .
\end{split}
\ee

whose solution reads
\be \label{solutionak}
\begin{split}
a_{k} \, & = - \, (1 \, + \, \fr{3k \, + \, 5}{2\, (k +1)\, (k + 2)} ) \, , \\
b_{k} \, & =  \, \fr{1}{2\, (k +1)\, (k + 2) \, (k + 3)} \, , \\
c_{k} \, & =  \, \fr{3}{2\, (k +1)} \, , \\
d_{k} \, & =  \,- \, \fr{3}{2\, (k +1)\, (k + 2) \, (k + 3)} \, . 
\end{split}
\ee

The evaluation of the traces of $\D$ is discussed in Appendix \ref{C}. Here
we only report the result:

\be
\begin{split}
\D^{\, [k]} \,  =  \, - \, 3 \, \sum_{q,\, l, \, m, \, t} \, \fr{a_{q\, l \, m}}{l} \,\{ 
& \a_{k,\, t} \, \pr^{\, l - 1 - 2 k + 2 t} \, \prd^{ \, m} \, \cF^{\, [q + t]}  + 
 \b_{k, \, t} \, \pr^{\, l - 2 k + 2 t} \, \prd^{ \, m + 1} \, \cF^{\, [q + t]} \\
& + \, \g_{k, \, t} \, \pr^{\, l + 1 - 2 k + 2 t} \, \prd^{ \, m + 2} \, \cF^{\, [q + t]} \, + \, irr \} \, ,
\end{split}
\ee

where the coefficients $\a_{k,\, t}$, $\b_{k,\, t}$ and $\g_{k,\, t}$ are given by the relations
%because of their symmetry property
%\be
%\begin{split}
%\a_{k,\, k - t} \, & = \, \a_{k,\, t} \, , \\
%\b_{k,\, k - t} \, & = \, \b_{k, \, t}\,  , \\
%\g_{k,\, k - t} \, & = \, \g_{k, \, t}\, ,
%\end{split}
%\ee
%are completely fixed by the values they assume for $t \, \leq \, [\fr{k + 1}{2}]$,
%which is
\be \label{solutiondelta}
\begin{split}
\a_{k,\, t} \, & = \, {k \choose t} \, , \\
\b_{k,\, t} \, & = \,2 \, (t + 1) \, {k \choose t + 1} \,  , \\
\g_{k,\, t} \, & = \, \,2 \, (t + 1) \, (t + 2) \, {k \choose t + 2} \, .
\end{split}
\ee

Eqs. (\ref{bzero2}), (\ref{solutionak}) and (\ref{solutiondelta}), together
with the coefficients (\ref{azero}), (\ref{auno}) and (\ref{adue}), and 
also together  with the relation

\be \label{generalfron}
\prd^{m} \, \cF^{\, [q]} \, = \, \fr{m \, (m - 1)}{2} \, \Box^{\, 2} \, \prd^{m - 2}
\, \vf^{\, [q + 1]} \, + \, [m \, (2q - 1) \, + \, (q + 1)]
\, \Box \, \prd^{m}\, \vf^{\, [q]} \, + \, irr \, ,
\ee  
giving the contribution to $\Box \vf^{\, [q]}$ and
$\Box \vf^{\, [q + 1]}$
from the m-th divergence and the q-th trace 
of the Fronsdal tensor, 
collectively represent the solution to our problem, and allow 
the explicit computation of the  contribution in $\Box \vf^{\, [k + 2]}$
contained in $\cB_0^{\, [k]}$. 

 By means of those formulas it is possible to check 
the conjecture up to any desired order (we did it up
to spin $11$). Nonetheless, it is still to be proved
that the sum of all the 
coefficients gives (\ref{tobeproven}) for an 
arbitrary value of $k$.

%%%%%%%%%%%%%%%%%%%%%%%%%%%%%%%%%%%%%%%%%%%%%%%%%%%%%%%%%%%%%%%%%%%%%
%%%%%%%%%%%%%%%%%%%%%%%%%%%%%%%%%%%%%%%%%%%%%%%%%%%%%%%%%%%%%%%%%%%%%
\scs{Conclusions}\label{conclusions}

 In this work we have proposed a Lagrangian description
of massive higher-spin fields, based on the massless, unconstrained
Lagrangians introduced in \cite{fs1, fs2, FMS}. The main 
results of the present analysis  are eqs. (\ref{mass}) and (\ref{massferm}), 
providing the generalised Fierz-Pauli mass terms for 
bosons and fermions respectively. 
To the best of our knowledge, this is the first
description of massive higher-spin theories 
which does not involve any auxiliary fields.

 The algebraic meaning of the generalised mass terms can be 
traced to the necessity to recover the Fierz-Pauli constraints
(\ref{fpagain}) and (\ref{fpferm}), in a context where  
neither auxiliary fields are introduced nor algebraic 
constraints are assumed, that might otherwise help in removing 
the lower-spin parts from the tensors used in the description.

 On the background of this work there is the general motivation
to try to investigate the possibility that
quantities amenable to a geometric interpretation
could play some meaningful role in the theory
of higher-spin fields. In order to keep as symmetric as possible 
the description of integer and half-integer spins, here we
also proposed an account of the fermionic theory
in which, as for the bosonic case, all quantities of 
dynamical interest can be defined in terms of curvatures.

 The issue of uniqueness, analysed in \cite{FMS}
for the massless theory, is met again in the present
treatment, since the generalised  Fierz-Pauli mass terms 
are such that \emph{any} geometric Lagrangian can be 
promoted to a consistent massive theory by means
of the same quadratic deformation. On the other hand, 
this appears to be more an algebraic consequence
of the strength of the constraints (\ref{fpagain}) and (\ref{fpferm})
rather than a deep issue. In our opinion indeed, as 
accounted in Section \ref{section4}, there
are indications that the mass term
(\ref{mass}) (and (\ref{massferm}), in a
possible extension of these arguments to fermions) actually
bears a direct relationship with the ``preferred'' 
geometric theory,  selected in \cite{FMS} on the basis 
of the requirement that the coupling with an external source be consistent.

  The main drawback of the use of curvatures for the description
of higher-spin dynamics is the presence of singularities in the 
Lagrangian, in the form of inverse powers of the D'Alembertian operator. 
These possible non-localities might be either the signal that there could be 
some intrinsic obstacle to such a geometric description, or, more
optimistically, the degenerate effect
at the linear level of some unusual feature of the full theory.

 About this point we can observe that, in the massive
equations described in this work, the presence of the singularities
plays a milder role. Actually, as can be better
appreciated by working in
Fourier transform, in the massive equations there are at least no poles
on the mass-shell and, in this sense, the generalised
Fierz-Pauli mass terms
(\ref{mass}) and (\ref{massferm}) provide
a kind of ``regularisation mechanism'' for the non-local, massless  theory.
 As the massless limit
is taken, the poles move towards the physical
region, but simultaneously the theory develops
a gauge symmetry, and once the limit is fully 
performed, all of them appear to be 
secluded in a pure-gauge sector of the equations of motion.

 Whether this should be taken as an indication that
the non-localities are actually harmless is far from obvious.
Nonetheless, a stronger argument is given by the possibility
of rephrasing all the properties of the geometric, non-local
theory in the local setting introduced in \cite{fs3, fsRev},
(and extended to (A)dS spaces in \cite{FMS}),
which results to be the minimal local setting where
the description of higher-spin fields in an unconstrained fashion 
is at all possible.

 In these local Lagrangians  the role of the singular
terms of the geometric description is replaced by
an auxiliary field, that acts as a ``compensator'',
to balance the gauge variation of the terms in the
equations of motion which are usually discarded
in the constrained, non-geometric approaches.
The ``memory'' of the singular nature of this field 
is in the presence of higher derivatives in its kinetic term
which again, notwithstanding the possibility
of fixing a gauge where this field vanishes, 
might be interpreted as the signal of something
odd in the general approach.

 As an answer to this objection, we have
proposed in this work a simple generalisation
of the local Lagrangians of \cite{fs3, fsRev} in which, 
at the price of introducing two more fields, for a total of five fields for
all spins, any higher-derivative term
disappears and the same dynamics of \cite{fs3, fsRev}, 
and then, from our viewpoint, the full geometrical
picture underlying the  linear theory, appears to be described 
in a completely conventional fashion.

%%%%%%%%%%%%%%%%%%%%%%%%%%%%%%%%%%%%%%%%%%%%%%%%%%%%%%%%%%%%%%%%%%%%%

\vskip 24pt

%%%%%%%%%%%%%%%%%%%%%%%%%%%%%%%%%%%%%%%%%%%%%%%%%%%%%%%%%%%%%%

\section*{Acknowledgments}

 It is a pleasure to thank X. Bekaert, J. Engquist, G. Ferretti and R. Marnelius
for helpful conversations, and especially J. Mourad and A. Sagnotti for 
stimulating discussions and collaboration. 
For the kind hospitality extended to me while part of this work was being done,
I am grateful to the APC-Paris VII and to the Scuola Normale Superiore
of Pisa, where the visit was supported in part by the MIUR-PRIN
contract 2003-023852. The present research was supported by the 
EU contract MRTN-CT-2004-512194.

%%%%%%%%%%%%%%%%%%%%%%%%%%%%%%%%%%%%%%%%%%%%%%%%%%%%%%%%%%%%%%

\newpage

\begin{appendix}

%%%%%%%%%%%%%%%%%%%%%%%%%%%%%%%%%%%%%%%%%%%%%%%%%%%%%%%%%%%%%%

\scs{Notation and conventions}\label{A}

%%%%%%%%%%%%%%%%%%%%%%%%%%%%%%%%%%%%%%%%%%%%%%%%%%%%%%%%%%%%%%

The space-time metric is the flat, mostly-positive one
in dimension $D$. If not otherwise specified, symmetrized indices are always left implicit.
In addition, traces are denoted by ``primes'' or 
by a number in square brackets: $\vf^{\, \prime}$ is thus the trace
of $\vf$, $\vf^{\, \prime\prime}$ is its double trace and 
$\vf^{\, [n]}$ is the $n$-th trace. 

 This notation results in an effective calculational procedure, 
whose basic rules are summarised in a number of identities, which reflect
some simple combinatorics. These rest on our convention of working
with symmetrized objects \emph{not} of unit strength, which is
convenient in this context but is not commonly used. For instance,
given the pair  of vectors $A_\mu$ and $B_\nu$, $A\, B$ here stands for
$A_\mu\, B_\nu + A_\nu\, B_\mu$, without additional factors of two.
The key identities are then:
\begin{alignat}{4}
&\left( \pr^{\, p} \, \vf  \right)^{\, \pe} \ \, & = & \ \, \Box \,
 \pr^{\, p-2} \, \vf \ + \, 2 \, \pr^{\, p-1} \,  \prd \vf \ + \, \pr^{\, p} \,
\vf^{\, \pe} \,  , \nonumber \\
& \partial^{\, p} \, \partial^{\, q} \ \, & =  &\ \, {p+q \choose p} % \binom{p+q}{p} \ \
\partial^{\, p+q} \, , 
\nonumber \\
&\partial \cdot  \left( \partial^{\, p} \ \vf \right) \ \, & = & \ \, \Box \
\partial^{\, p-1} \ \vf \ + \
\partial^{\, p} \ \partial \cdot \vf \, ,  \nonumber \\
& \partial \cdot  \eta^{\, k} \ \, & = & \ \, \partial \, \eta^{\, k-1} \, , \\
&\left( \eta^k \, \vf  \,  \right)^{\, \prime} \, \ & = & \ \, \left[ \, D
\, + \, 2\, (s+k-1) \,  \right]\, \eta^{\, k-1} \, \vf \ + \ \eta^k
\, \vf^{\, \prime} \, , \nonumber \\
&(\vf \, \psi)^{\, \pe} \ \, & =& \ \, \vf^{\, \pe} \, \psi \, + \, \vf \, \psi^{\, \pe} \, + \, 2
\, \vf \cdot \psi \, , \nonumber \\
& \eta \, \eta^{\, n-1} \ \, & = & \  \, n \, \eta^{\, n} \, , \nonumber \\
&\g \, \cdot \, (\g \, \psi) \ \, & = & \  \, (D \, + \, 2 \, s) \, \psi \, - \, \g \, \psisl \, , \nonumber
\end{alignat}
where in particular in the last equality $\psi$ is a rank-$s$ spinor-tensor.
As anticipated, the basic ingredient in these expressions is the
combinatorics, which is simply determined by the number of relevant
types of terms on the two sides. Thus, for a pair of flat
derivatives, $\partial \,
\partial = 2 \, \partial^{\, 2}$ reflects the fact that, as a result of
their commuting nature, the usual symmetrization is redundant
precisely by the overall factor of two that would follow from the
second relation. In a similar fashion, for instance, the
identity $\h \, \h^{\, n-1} = n \, \h^{\, n}$ reflects the different numbers of terms generated by the
naive total symmetrization of the two sides: $\left( {2 n} \atop {2}
\right) \times (2 n -1)!!$ for the expression on the \emph{l.h.s},
and $(2 n + 1)!!$ for the expression on the \emph{r.h.s.}.

%%%%%%%%%%%%%%%%%%%%%%%%%%%%%%%%%%%%%%%%%%%%%%%%%%%%%%%%%%%%%%

\scs{On the explicit form of $\cA_{\, \vf}$}\label{B}

%%%%%%%%%%%%%%%%%%%%%%%%%%%%%%%%%%%%%%%%%%%%%%%%%%%%%%%%%%%%%%

The form of $\g_{\, \vf}$ for the first cases of spin $s\, = \, 3, 4, 5, 6$
\be \label{explicitgammas}
\begin{split} 
\g_{\, 3} \, & = \,  \fr{1}{3\, \Box^{\, 2}} \prd \cF^{\, \pe}  \, , \\
\g_{\, 4} \, & = \,  \g_{\, 3} \, - \, 
\fr{1}{3}\, \fr{\pr}{\Box^{\, 3}} \prd \prd \cF^{\, \pe} \, + \, 
\fr{1}{12}\, \fr{\pr}{\Box^{\, 2}}\,\cF^{\, \pe \pe} \, , \\
\g_{\, 5} \, & = \,   \g_{\, 4} \, + \,  \fr{2}{5}\, \fr{\pr^{\, 2}}{\Box^{\, 4}}\, 
\prd \prd \prd \cF^{\, \pe} \, -\, 
\fr{1}{5}\, \fr{\pr^{\, 2}}{\Box^{\, 3}}\prd \cF^{\, \pe \pe} \, , \\
\g_{\, 6} \, & = \, \g_{\, 5} \, - \, 
\fr{8}{15}\, \fr{\pr^{\, 3}}{\Box^{\, 5}}\,\prd^{4} \cF^{\, \pe}\, + \,
\fr{2}{5}\, \fr{\pr^{\, 3}}{\Box^{\, 4}}\prd \prd \cF^{\, \pe \pe} \, - \, 
\fr{1}{30}\, \fr{\pr^{\, 3}}{\Box^{\, 3}} \, \cF^{\, \pe \pe \pe} \, ,
\end{split}
\ee
suggests the recursive relation (whose general
validity was proven in \cite{FMS})\ft{Since in this Section we 
want to keep track of the spin $s$, we switch to a more
explicit notation, renaming $\g_{\, \vf} \ \ \ra \ \  \g_{\, s}$.
Remember that $\g_{\, s}$, referring to the case of spin $s$, is 
a rank-$(s-3)$ tensor.}
 \be
\g_{\, s} \, = \, \g_{\, s-1} \, + \, \pr^{\, s-3} \, 
\{a_1 \,\fr{1}{\Box^{\, s-1}}\, \prd^{s-2} \cF^{\, \pe} \, + \,
\dots \, + \, a_k \, \fr{1}{\Box^{\, s-k}} \, \prd^{s-2k} \cF^{\, [k]} \, + \, 
\dots \, + \, a_n \, 
\begin{cases}
\fr{1}{\Box^{\, n}}\,  \cF^{\, [n]}        \, ,\\
\fr{1}{\Box^{\, n + 1}}\, \prd \cF^{\, [n]}  \, ,
\end{cases}\}
\ee
where the last two options refers to the cases of spin $s = 2n$ and $s = 2n + 1$, respectively.
The crucial point is that \emph{one
coefficient, regardless of the spin, is uniquely associated to each structure\ft{Here, as in the 
rest of this Appendix, for the sake of brevity we shall simply
call ``structure'' a generic term in $\g_{\, s}$ of the form
$\pr^{\, l}\, \prd^{\, m} \, \cF^{\, [k]}$.}.}  It is also important that in every term
in $\g_s \, - \, \g_{s-1}$ the Fronsdal tensor $\cF$ is fully saturated.

 The general form of $\g_s$, 
with a more appropriate definition of the coefficients,
can be written
\be \label{structures}
\g_s \, = \, \sum_{k, l, m} \, a_{k l m} \, \fr{\pr^{\, l}}{\Box^{\, m + k}} \, \prd^m \, \cF^{\, [k]} \, ,
\ee
where
\be \label{rangeklm}
\begin{split}
k \, &= \, 1 \, , \dots \, [\fr{s}{2}] \, , \\
l \, &= \, 0\, ,\dots, s-3 \, , \\
m\, &= \, l \, + \, 3 \, - \, 2 \, k \, \\
m \,&\geq \, 0 \, . 
\end{split}
\ee
 We would like to find $a_{k l m}$ for generic 
values of $k, l, m$.
  
 The idea is to exploit the identities (\ref{Aident}) satisfied by  $\cA_{\, \vf}$.
As we shall see, they 
play quite different roles, the main reason being that, in order to 
deduce from those conditions  equations for the coefficients
$a_{k l m}$, one has to make sure that there are no cancellations
between terms in the field $\vf$ contained in different structures. This is true 
for the double-tracelessness condition, but it is \emph{not}
true for the Bianchi identity, but for one single exception, 
that will be anyway important.

 Let us also observe that the general strategy will be to consider 
$\g_{s-1}$ as known, and to try and determine all coefficients for the 
structures in $\g_s \, - \, \g_{s-1}$. This implies for example that 
our unknowns will always be of the form $a_{k, s-3, s - 2k}$.

\subsubsection*{Double-tracelessness}
The condition
\be
\cA_{\, \vf}^{\, \pe \pe} \, \eq \, 0 \, ,
\ee
is equivalent to 
\be \label{doubletrace}
(\pr^{\, 3} \, \g_s)^{\, \pe \pe} \,\eq \,  \fr{1}{3} \, \cF^{\, \pe \pe} \, ,
\ee
where 
\be \label{gammas}
(\pr^{\, 3} \, \g_s)^{\, \pe \pe} \, = \, 4\, \Box \, \prd \g_s \, + \, 
4 \, \pr \, \prd \prd \g_s \, + \, 4 \, \pr^{\, 2} \, \prd \g_s^{\, \pe} \, 
+ \, 2 \, \Box \, \pr \, \g_s^{\, \pe} \, + \, \pr^{\, 3} \, \g_s^{\, \pe \pe} \, .
\ee
Computing each term in (\ref{gammas}) by means of (\ref{structures})
and imposing (\ref{doubletrace}) leads to 
the system\ft{Since the coefficients of each structure in $\g_{\, s}$
are spin-independent, and (\ref{doubletrace}) must hold for 
any value of $s$, the only possibility is that in this relation the coefficients of the
various structures all vanish.}
\be
\begin{split}
& {s \choose 3} a_{k, s-3, s-2k} \, + \, 
4 \, {s - 1 \choose 3} a_{k, s - 4, s- 2k - 1} \, + \,
4\, {s  - 2 \choose 3} a_{k, s - 5, s - 2k - 2} \, + \, \\
& 2\, {s  - 2 \choose 3} a_{k - 1, s - 5, s - 2k} \, + \,
4\, {s  - 3 \choose 3} a_{k - 1, s - 6, s - 2k - 1} \, + \,
4\, {s  - 4 \choose 3} a_{k - 2, s - 7, s - 2k} \, = \, 0 \, .
\end{split}
\ee
 Let us notice that in each term we have one independent variable ($k$) and 
one parameter ($s$); keeping this in mind, we can simplify the notation
defining
\be
{s \choose 3} \, a_{t, s - 3, m} \, \eq \, b_{s, m} \, ,
\ee
which is free of ambiguities, since for a given $s$ there is
a one-to-one relation between $t$ and $m$.  
In this notation the system can be written in the more compact form
\be \label{systemtraces}
b_{s, m} \, + \, 2 \, b_{s - 2, m} \, + \, b_{s - 4, m} \, = \, 
- \, 4 \, \{b_{s - 1, m - 1} \, + \, b_{s - 2, m - 2} \, + \, b_{s - 3, m - 1} \} \, .
\ee
In particular  the system written as in (\ref{systemtraces}) is ready 
to be solved `by traces', keeping the number of divergences fixed.
Of course, it would also be possible to keep the number of traces fixed and to
solve the system `by divergences'. 

There are two kinds of difficulties:
\begin{description}
 \item[$\ra$] the need for an infinite list of initial conditions,
 \item[$\ra$] the iteration of the r.h.s. of (\ref{systemtraces}).
\end{description}

To begin with, we can solve a few cases by making use of the list
(\ref{explicitgammas}) to get the initial conditions. We find in this 
way the coefficients referring to structures
involving zero, one and two 
divergences\ft{Here the label ``$s$'' refers to the value
of the spin in correspondence of which the structure multiplied by $a_{k l m}$
first appears. We would like to stress once again that every structure appears for the first time
for a value of the spin such that the Fronsdal tensor is completely saturated. This implies 
for example that for $m = 0$ we must saturate all indices by traces, and then the spin \emph{must}
be even, when the corresponding structure first appears. Similar reasonings
explain why for $m = 1$ and $m = 2$ there is no need to take the 
integer parts of $\fr{s - 1}{2}$ and $\fr{s - 2}{2}$ respectively.}:
\be \label{azero}
a_{\fr{s}{2}, \, s - 3, \, 0} \, = \, \fr{(-\, 1)^{\fr{s}{2}}}{s\, (s \, - \, 1)} \, ,
\ee
\be \label{auno}
a_{\fr{s - 1}{2},\, s - 3,\, 1} \, = \, \fr{(-\, 1)^{\fr{s + 1}{2}}}{s} \, ,
\ee
\be \label{adue}
a_{\fr{s - 2}{2},\, s - 3,\, 2} \, = \, (-\, 1)^{\fr{s}{2} + 1}\fr{s\, - \, 2}{2\, (s \, - \, 1)} \, .
\ee
These coefficients will be the ones actually needed to check
the conjecture presented in Section \ref{section4}.
As a matter of principle, if we want the coefficient for a structure with a certain 
\emph{fixed} (as a number, not as a parameter) number of divergences, 
we could compute explicitly, by hand, the 
$\g_s$ where this structure first appears, and use this initial condition 
to solve the system in general, finding the corresponding solution for 
any (allowed) $s$. In this sense, at the price of performing
\emph{one} explicit computation up to a certain value of the spin, 
we are then in the position to find \emph{a class} of coefficients, as for the cases
displayed above. 

 On the other hand, at least to allow in principle the 
possibility of finding the general solution in closed form,
we have to ask ourselves whether
we can find an infinite list of initial data. This is the point where the
Bianchi identity helps.

\subsubsection*{Bianchi identity}

From the first of (\ref{Aident}) 
\be \label{Aidentapp}
\prd \cA_{\, \vf} \, - \, \12 \, \pr \, \cA_{\, \vf}^{\, \pe} \, \eq \, 0 \, ,
\ee
we find the following condition on $\g_s$
\be \label{bident2}
\vf^{\, \pe \pe} \, = \, 4 \, \prd \g_s \, + \, \pr \, \g_s^{\, \pe} \, ,
\ee
which can be regarded as the solution to the problem 
of inverting $\cF^{\, \pe} \, (\vf, \vf^{\, \pe}, \vf^{\, \pe \pe})$ w.r.t. $\vf^{\, \pe \pe}$.
In order to better explain its meaning, we 
would like to stress two points:
\begin{description}
 \item[$\ra$] the identity (\ref{bident2}), that could naively look in some 
sense expected (because of the gauge transformation of $\vf^{\, \pe \pe}$), actually
represents a non trivial relation, given that, as stressed in the footnote
at page \pageref{notagamma}, there are infinitely many
other $\g$'s such that $\d \, \g \, = \, \L^{\, \pe}$, none of which 
would satisfy it.
 \item[$\ra$] Moreover,  (\ref{bident2}) makes it 
manifest that in the Bianchi identity cancellations 
must occur \emph{among} different structures; indeed, the 
very presence of a term in the naked double trace of $\vf$
will necessarily imply a chain of compensations among the various structures.
\end{description}  
This last observation can be 
easily made more concrete via a specific example.
Consider the case of spin $6$, and use the form of $\g_6$ 
given in (\ref{explicitgammas}) to compute 
the two contributions to (\ref{bident2}):
\be
\begin{split}
4 \, \prd \g_6 \, &= \, \fr{4}{15} \, \pr \, \prd^3 \cF^{\, \pe} \, 
- \, \fr{8}{15} \, \pr^{\, 2} \, \prd^4 \cF^{\, \pe} \, + \, \fr{1}{3} \, \cF^{\, \pe \pe} \, \\
 & \, - \, \fr{7}{15} \, \pr \, \prd \cF^{\, \pe \pe} \, + \, 
\fr{4}{5} \, \pr^{\, 2} \, \prd^2 \cF^{\, \pe \pe} \, - \, 
\fr{2}{15} \, \pr^{\, 2} \, \cF^{\, \pe \pe \pe} \, , \\
\pr \, \g_6^{\, \pe} \, & =\, - \, \fr{4}{15} \, \pr \, \prd^3 \cF^{\, \pe} \, 
+ \, \fr{8}{15} \, \pr^{\, 2} \, \prd^4 \cF^{\, \pe} \, + \,
+\, \fr{3}{10} \, \pr \, \prd \cF^{\, \pe \pe} \, \\
&\,  - \, \fr{2}{3} \, \pr^{\, 2} \, \prd^2 \cF^{\, \pe \pe} \, + \, 
\fr{1}{10} \, \pr^{\, 2} \, \cF^{\, \pe \pe \pe} \, .
\end{split}
\ee
We can see that in general the single structures in (\ref{bident2}) will not have 
vanishing coefficients. For example
the term $\fr{1}{3} \, \cF^{\, \pe \pe}$ in $4 \, \prd \g_6$ contains the only 
term in $\vf^{\, \pe \pe}$ ``naked'' of the whole r.h.s. of
(\ref{bident2}), to be matched with the l.h.s. of the same expression. 
On the other hand it also contains contributions in $\prd \vf^{\, \pe \pe}$ and 
$\vf^{\, \pe \pe \pe}$ that will cancel because of analogous terms contained 
in other structures. For this reason, (\ref{Aidentapp}) could not
have been used to derive a system like (\ref{systemtraces}).

 It is then remarkable that \emph{structures involving \emph{one} trace of $\cF$ 
exactly compensate} each other. This is a general result, and it is due to the fact that
a structure of the form $\prd^m \cF^{\, \pe}$ contains a contribution in 
$\prd^{m + 2} \vf$ that \emph{cannot} be compensated by anything else in 
(\ref{bident2}), given that $\cF^{\, [k]}$ contains at least $\vf^{[k - 1]}$, 
$\forall k \geq 2$.
This means that it is possible to use the Bianchi identity to find an equation
for structures involving one trace of $\cF$, together with an 
arbitrary number of divergences. The solutions to this equation would 
provide exactly the list of initial conditions we need for 
the system (\ref{systemtraces}).
 We determine in this way a new class of coefficients:
\be
 a_{1, s - 3, s - 2} \, = \, (-1)^{s + 1} \, \fr{2^{\, s}}{4 \, s \, (s \, - \, 1)} \, ,
\ee
allowing in principle to find the general solution to
(\ref{systemtraces}), for an arbitrary number of traces and divergences
of the corresponding structure.  Even if such a general solution so far is not known, 
the coefficients explicitly determined are the only ones needed for 
the discussion of the relation between the geometric Einstein tensor
(\ref{einstein}) and the generalised Fierz-Pauli mass term (\ref{mass}).

%%%%%%%%%%%%%%%%%%%%%%%%%%%%%%%%%%%%%%%%%%%%%%%%%%%%%%%%%%%%%%

\scs{Evaluation of $\D^{\, [k]}$} \label{C}

%%%%%%%%%%%%%%%%%%%%%%%%%%%%%%%%%%%%%%%%%%%%%%%%%%%%%%%%%%%%%%

  Starting from the general form of $\D$, given in (\ref{delta})
 
\be %\label{delta}
\D \, = \, - \, 3 \, \sum_{q, \, l, \, m} \, \fr{a_{q l m}}{l} \, 
\pr^{\, l - 1} \, \prd^{ \, m} \, \cF^{\, [q]} \, ,
\ee
the computations
of the $k$-th trace will generate contributions containing, 
for a given coefficient $a_{q l m}$, all powers of gradients
from $l-1$ to $l-1-2k$. We have to take into account the general
result (\ref{generalfron}) for an arbitrary trace and an arbitrary divergence
taken on the Fronsdal tensor, that we report here for simplicity
\be %\label{generalfron}
\prd^{m} \, \cF^{\, [q]} \, = \, \fr{m \, (m - 1)}{2} \, \Box^{\, 2} \, \prd^{m - 2}
\, \vf^{\, [k + 1]} \, + \, [m \, (2q - 1) \, + \, (q + 1)]
\, \Box \, \prd^{m}\, \vf^{\, [q]} \, + \, irr \, ,
\ee  
showing that, for our present purposes, all multiple
divergences of $\cF$ with $m \geq 3$ can be classified as irrelevant, 
since they will contain at least one divergence of $\vf$ and thus cannot contribute
to $\Box \vf^{\, [k + 2]}$ in (\ref{tobeproven}). Under this condition, 
it is possible to observe that the various contribution to $\D^{\, [k]}$
for a given $a_{q l m}$ display a ``reflection symmetry'', as can 
be appreciated for instance in the explicit example of $\D^{\, [4]}$
\be %\label{delta}
\begin{split}
\D^{\, [4]} \, = \, - \, 3 \, & \sum_{q, \, l, \, m}  \, \fr{a_{q l m}}{l} \, \{
 \pr^{\, l - 9} \, \prd^{ \, m} \, \cF^{\, [q]} \, + 
\, 8 \, \pr^{\, l - 8} \, \prd^{ \, m + 1} \, \cF^{\, [q]} \,+
\, 4 \, \pr^{\, l - 7} \, \prd^{ \, m} \, \cF^{\, [q + 1]} \,+ \\
&24 \,  \pr^{\, l - 7} \, \prd^{ \, m + 2} \, \cF^{\, [q]} \, + 
\, 24 \, \pr^{\, l - 6} \, \prd^{ \, m + 1} \, \cF^{\, [q + 1]} \,+
\, 6 \, \pr^{\, l - 5} \, \prd^{ \, m} \, \cF^{\, [q + 2]} \,+ \\
&48 \,  \pr^{\, l - 5} \, \prd^{ \, m + 2} \, \cF^{\, [q + 1]} \, + 
\, 24 \, \pr^{\, l - 4} \, \prd^{ \, m + 1} \, \cF^{\, [q + 2]} \,+
\, 4 \, \pr^{\, l - 3} \, \prd^{ \, m} \, \cF^{\, [q + 3]} \,+ \\
&24 \,  \pr^{\, l - 3} \, \prd^{ \, m + 2} \, \cF^{\, [q + 2]} \, + 
\, 8 \, \pr^{\, l - 2} \, \prd^{ \, m + 1} \, \cF^{\, [q + 3]} \,+
 \, \pr^{\, l - 1} \, \prd^{ \, m} \, \cF^{\, [q + 4]} \} \, + \, irr \\
\end{split}
\ee
in the sense that, for a fixed number of divergences, there
is coincidence between the coefficient of the term in 
$\cF^{\, [q + t]}$ and the coefficient of $\cF^{\, [q + 4 - t]}$.
Because of this symmetry we can limit ourselves to the 
computation of the coefficients for $t \leq [\fr{k + 1}{2}]$.
To clarify the notation, we introduce three kinds of coefficients, 
according to the following table:
\be
\begin{split}
\a_{k,\, t} \, & \  \ra \  \mbox{terms with no extra divergences, and $t$
extra traces} \  \ra \  \prd^{\, m} \, \cF^{\, [q + t]}  \,  , \\
\b_{k, \, t}\, & \  \ra \  \mbox{terms with one extra divergence, and $t$
extra traces} \  \ra \  \prd^{\, m + 1} \, \cF^{\, [q + t]}  \,  , \\
\g_{k, \, t}\, & \  \ra \  \mbox{terms with two extra divergences, and $t$
extra traces} \  \ra \  \prd^{\, m + 2} \, \cF^{\, [q + t]}  \,  , 
\end{split}
\ee
and write $\D^{\, [k]}$ according to the formula
\be
\begin{split}
\D^{\, [k]} \,  =  \, - \, 3 \, \sum_{q,\, l, \, m, \, t} \, \fr{a_{q\, l \, m}}{l} \,\{ 
& \a_{k,\, t} \, \pr^{\, l - 1 - 2 k + 2 t} \, \prd^{ \, m} \, \cF^{\, [q + t]} \, + \\
&\b_{k, \, t} \, \pr^{\, l - 2 k + 2 t} \, \prd^{ \, m + 1} \, \cF^{\, [q + t]} \, + \\
& \g_{k, \, t} \, \pr^{\, l + 1 - 2 k + 2 t} \, \prd^{ \, m + 2} \, \cF^{\, [q + t]} \, + \, irr \} \, ,
\end{split}
\ee
where
\be %\label{rangeklm}
\begin{split}
q \, &= \, 1 \, , \dots \, [\fr{s}{2}] \, , \\
l \, &= \, 1\, ,\dots, s-3 \, , \\
m\, &= \, l \, + \, 3 \, - \, 2 \, q \, , \\
t \,  &= \, 0 \, \dots \, k \, . 
\end{split}
\ee
It is then possible to show that the coefficients 
satisfy the recursive system
\be
\begin{split}
 \a_{k,\, t}  \, & = \,  \a_{k - 1,\, t} \, + \,  \a_{k - 1,\, t + 1} \, ,  \\
 \b_{k,\, t}  \, & = \,  \b_{k - 1,\, t} \, + \,  \b_{k - 1,\, t - 1} \, + \, 2 \, \a_{k - 1,\, t} \, ,  \\
 \g_{k,\, t}  \, & = \,  \g_{k - 1,\, t} \, + \,  \g_{k - 1,\, t - 1} \, + \, 2 \, \b_{k - 1,\, t} \, ,  
\end{split}
\ee
whose solution is %for $t \leq [\fr{k + 1}{2}]$,
\be
\begin{split}
 \a_{k,\, t}  \, & = \, \fr{1}{t !} \, k \, (k - 1) \, \dots \, (k - t + 1 ) \, = \, {k \choose t} \, ,  \\
 \b_{k,\, t}  \, & = \, \fr{2}{t !} \, k \, (k - 1) \, \dots \, (k - t + 1 ) \, (k - t) \, = \,2 \,(t + 1) {k \choose t + 1} \, ,  \\
 \g_{k,\, t}  \, & = \, \fr{2}{t !} \, k \, (k - 1) \, \dots \, (k - t) \, (k - t - 1) \, = 
 \,2 \,(t + 1)\, (t + 2) \, {k \choose t + 2} \, .  \\
\end{split}
\ee

\end{appendix}

\newpage

%%%%%%%%%%%%%%%%%%%%%%%%%%%%%%%%%%%%%%%%%%%%%%%%%%%%%%%%%%%%%%%%%%%%%

\end{document}